\documentclass[superscriptaddress,twocolumn,preprintnumbers,amsmath,amssymb,nolongbibliography]{revtex4-2}

\usepackage[version=3]{mhchem} % Formula subscripts using \ce{}
\usepackage[dvipsnames]{xcolor}

\usepackage{graphicx}
\usepackage{dcolumn}
\usepackage{bm}
\usepackage{amsmath}
\usepackage{epsfig}
\usepackage{rotating}
\usepackage{enumerate}
\usepackage{amsfonts}
\usepackage{mathtools}
\usepackage{amssymb}
\usepackage[colorlinks]{hyperref}
\usepackage{siunitx}

\newcommand\myshade{80}
\colorlet{mylinkcolor}{ForestGreen}
\colorlet{mycitecolor}{Red}
\colorlet{myurlcolor}{violet}

\hypersetup{
  linkcolor  = mylinkcolor!\myshade!black,
  citecolor  = mycitecolor!\myshade!black,
  urlcolor   = myurlcolor!\myshade!black,
  colorlinks = true
}

\def\al2o3{{Al$_2$O$_3$}}

\NewDocumentCommand\shell{m}{\texttt{#1}}

\begin{document}

\title{Phonon dynamics for light dark matter detection}

\author{Mart\'i Raya-Moreno}
\affiliation{Institut de Ci\`encia de Materials de Barcelona, ICMAB-CSIC, Campus UAB, 08193 Bellaterra, Spain}

\author{Bradley J. Kavanagh}
\affiliation{Instituto de F\'isica de Cantabria (IFCA), University of Cantabria (UC)-CSIC,
Avenida de los Castros, s/n E-39005 Santander, Spain}
% \email{kavanagh@ifca.unican.es}

\author{Lourdes F\`abrega}
\affiliation{Institut de Ci\`encia de Materials de Barcelona, ICMAB-CSIC, Campus UAB, 08193 Bellaterra, Spain}
\email{lourdes@icmab.es}

\author{Riccardo Rurali}
\affiliation{Institut de Ci\`encia de Materials de Barcelona, ICMAB-CSIC, Campus UAB, 08193 Bellaterra, Spain}
\email{rrurali@icmab.es}

\date{\today}

\begin{abstract}
The search for low-mass dark matter (DM), which has gained much interest recently, goes in parallel with the identification of new detection channels and the development of suitable detectors. Detection of the resulting small energy depositions is challenging: it requires extremely high sensitivity, only achievable by cryogenic thermal detectors, which might be put to the limit. Understanding the processes which can limit performances of these detectors can be thus crucial for evaluating the feasibility of the proposed new detection schemes and, eventually, to design the detectors and tune their performance. 
In this paper we focus on a promising detection scheme, the excitation of single optical phonons in polar materials, to evaluate one of the possible --and less understood--  limiting factors of cryogenic thermal detectors, i.e.\ the phonon dynamics in the target/absorber. We present a detailed theoretical analysis, within an entirely {\it ab initio} scheme, of the downconversion and propagation processes undergone by optical phonons, created by the interaction of a low-mass DM particle in an \al2o3\ target, until they reach the interface with a phonon Al collector. After a preliminary methodological survey that reveals the limitations of  any Relaxation Time Approximation (RTA) based method, we developed a 3D beyond-RTA phonon Monte Carlo that allowed us to introduce the spatial dimension of the device and address questions about impact of target size and scattering position. We analyse also the effect of the phonon energy and wavevector. We also show that isotopes can, perhaps counterintuitively, result in a larger heat flux by providing transport channels of higher velocities, thus favoring detection. Our results suggest that, though challenging, the direct detection of light DM via athermal phonon generation appears feasible, and that the phonon downconversion followed by quasi-ballistic propagation does not appear to be a major bottleneck in terms of reducing the signal.
\end{abstract}

\maketitle

\section{Introduction}
\label{sec:intro}

The nature of DM, constituting most of the mass in our Universe, remains an elusive issue~\cite{Bertone:2004pz}. Through more than three decades, the most promising candidates have been considered to be Weakly Interacting Massive Particles (WIMPs), with masses in the GeV-to-TeV range. Yet, after decades of searching, no conclusive detection of WIMPs has been made~\cite{Gaskins:2016cha,Boveia:2018yeb,Billard:2021uyg}. As a consequence, the search for DM has diversified substantially in the last decade~\cite{Bertone:2018krk}, with lighter DM particles gaining much attention, both theoretically and experimentally~\cite{Essig:2011nj}.

The direct detection of sub-GeV DM particles poses major experimental challenges. Standard searches for DM-nucleus scattering become ineffective at low mass, as the low DM velocity in the Milky Way Halo ($\mathcal{O}(10^{-3})\,c)$)~\cite{Green:2017odb} and kinematic mismatch between the DM and nuclear masses limit the amount of energy which can be deposited. In recent years, several detection channels for light and ultralight DM have been theoretically proposed, such as the excitation of phonons, polaritons or magnons~\cite{KnapenPLB18,ChigusaPRD20,MitridatePRD20} or the absorption of DM in semiconductors~\cite{HochbergPRD17,BlochJHEP17} and superconductors~\cite{Hochberg:2016ajh,Hochberg:2019cyy}. In general, collective excitations must be taken into account when the de Broglie wavelength becomes larger that the particle spacing; when dealing with solid targets for DM detection, this corresponds roughly to particle masses below 1~MeV, becoming progressively relevant as the particle mass decreases. Furthermore, collective excitations can additionally imply a daily modulation in the signal if anisotropic targets are used~\cite{GriffinPRD18}. This could provide further evidence for DM, allowing us to more easily distinguish a genuine DM signal from unmodulated background noise.

\begin{figure*}[t]
\includegraphics[width=0.8\linewidth]{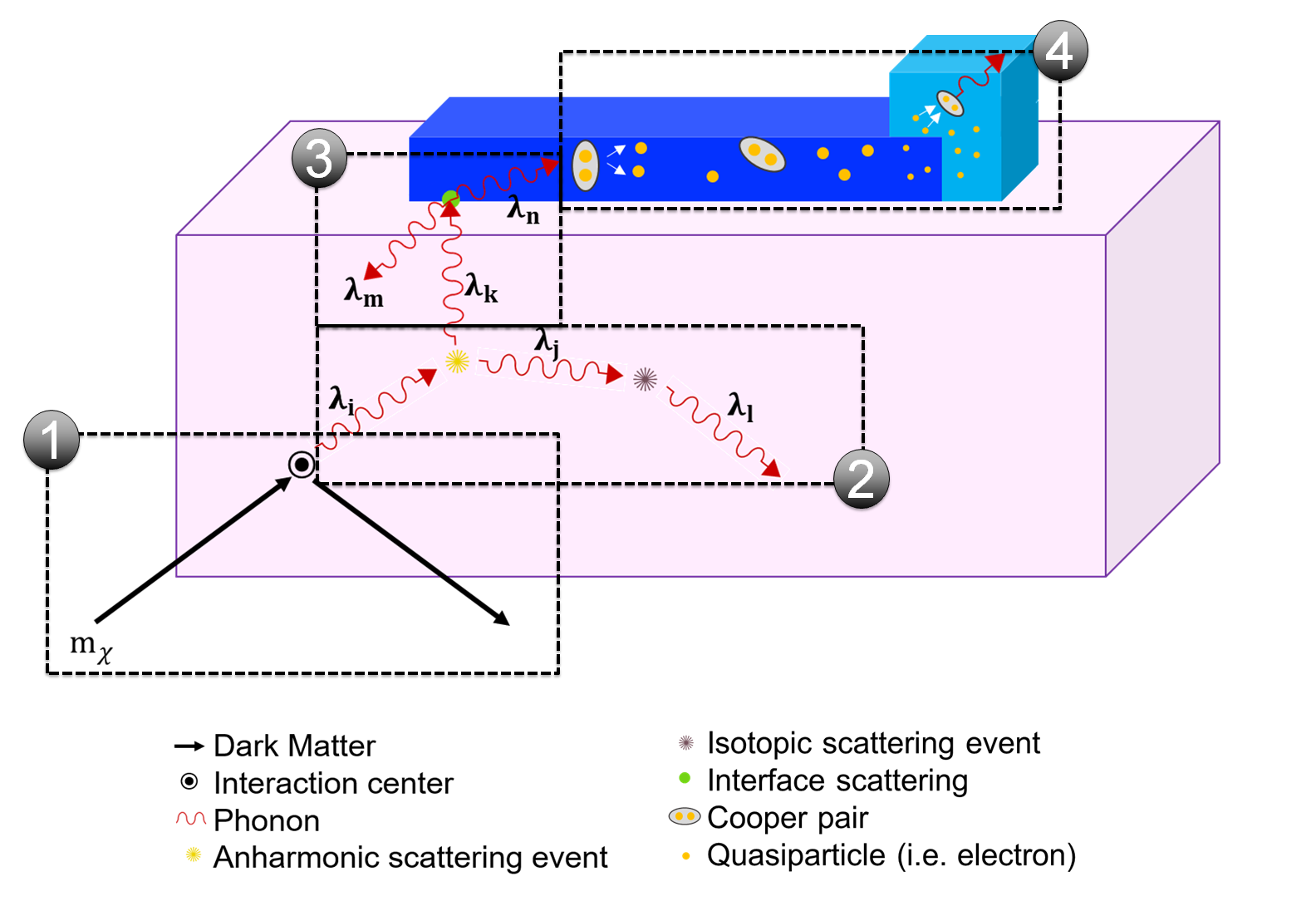}
       \caption{Scheme of a phonon-based DM detector  and of the          relevant physical processes involved.
                The DM particle generates phonons in the semiconductor, the target (1); these first suffer downconversion, travel until they reach the target backsurface (2) and are then transmitted to a superconducting thin film, the collector (3). Once inside the collector, phonons break Cooper pairs, generating quasi-particles (4). At the end of the phonon collector lies a Transition-Edge Sensor (TES).}
\label{fig:detector}
\end{figure*}

Among newly proposed strategies for the direct detection of low mass DM, of special interest is the excitation of optical phonons in polar materials~\cite{KnapenPLB18}, because of their significant coupling to light dark photon mediators. Moreover, optical phonons in polar materials display gaps of the order of 10-30~meV, favourable for the scattering of DM particles with masses below 1~MeV. The detection of such low energy excitations requires extremely high sensitivity, only achievable by cryogenic detectors. 
Indeed, cryogenic thermal sensors, operating at temperatures well below 1K and using a phonon signal to determine the total energy released in an absorber/target, have become essential to build up very sensitive rare event detectors for fundamental physics experiments and astrophysics~\cite{PirroARNPS17}. Transition Edge Sensors (TES)~\cite{IrwinTAP07} are widely used in or proposed for telescopes both on Earth and in space, to detect photons over a broad range of the electromagnetic spectrum. TES have also been used for years - or are being proposed - for particle detectors, such as those for dark matter (DM) searches (CRESST~\cite{AngloherAP05,StraussNIMA17} , SuperCDMS~\cite{AgnesePRL14,AgnesePRD17}, COSINUS~\cite{COSINUS:2023kqd}) or neutrino physics (HOLMES~\cite{DeGerone:2022dxb}, PTOLEMY~\cite{PTOLEMY:2018jst}). Other types of cryogenic detectors are also used or in development, in a variety of initiatives: NTDs (EDELWEISS~\cite{EDELWEISS:2017uga}); kinetic inductance detectors (KIDs)  (BULLKID~\cite{Cruciani:2022mbb}, CADEx~\cite{Aja:2022csb}, CALDER~\cite{Casali2019}); superconducting single photon nanowires for light DM~\cite{Chiles:2021gxk}; or magnetic microcalorimeters for neutrinos (ECHo~\cite{Mantegazzini:2023igy}).  Many cryogenic particle detectors are based on the so-called quasi-particle trap-assisted TES, QET~\cite{IrwinRSI95} (Figure~\ref{fig:detector}), in which a superconducting film, placed between the target and the TES, is used as a phonon collector. Its role is to decouple the phonon collection efficiency (related to the target surface area covering) from the TES sensitivity (proportional to TES volume), thus allowing independent optimization of both.

The detection of single phonons resulting from sub-MeV DM scattering will require fine tuning of detector performances (see e.g. Fink and coworkers~\cite{FinkAIPAdv20}), involving understanding and optimization of all the steps of the detection process. While the reach and daily modulation of the signal are mostly determined by the harmonic properties of the target material, the ultimate performances of the detector, in terms of efficiency and phonon energy sensitivity, are limited by a number of factors. These include the TES sensitivity and time response ($t_r \gtrsim \mu\mathrm{s}$), as well as the processes undergone by the phonons initially generated in the target as a result of the DM interaction (sometimes referred to as {\it athermal} phonons, to stress that they are strongly out-of-equilibrium quasi-particles that do not correspond to the equilibrium thermodynamic conditions of the target material).
% These processes are poorly understood and involve downconversion to long wavelength (quasi-ballistic) phonons, transport to the target surface and transmission to the phonon collector, through the target/collector interface (see the sketch in Figure~\ref{fig:detector}).
This means that knowing the phonon properties of the target (spectra, transport, role of defects, interface transmission rate) is crucial for designing and building a detector capable of detecting low mass dark matter through single phonon excitations. 

A benchmarking of polar materials as possible targets has been published recently by Griffin and coworkers~\cite{GriffinPRD20}, taking into account different detection channels.  There, the expected detector reach has been evaluated considering several types of DM candidates and using density-functional theory (DFT) calculations of phonon properties. This initial work helps in identifying the material parameters that determine the detector reach, and to make a pre-selection of target candidates.
Calculations at the harmonic level (i.e.\ phonon band structure) are indeed very useful as a pre-screening tool, but other relevant features can only be addressed by means of anharmonic calculations, where the cross-section of the relevant scattering processes are computed~\cite{FugalloPS18}.  In particular, the mechanisms for phonon downconversion, the conditions for the subsequent ballistic transport of long wavelength phonons and the role of isotope purity cannot be tackled at the harmonic level and have thus far been neglected. Only a recent paper~\cite{MartinezPRAppl19} explores the phonon transport in Si with GEANT~\cite{geant}, a software tool used in high energy physics to study particle-matter interactions. Although they provide some useful indications, these simulations are based on a number of oversimplifying assumptions that call into question the predictive power of these results. Similarly, phonon reflection at the interface with the superconducting collector is often neglected or treated within simplified phenomenological models.
% \bjk{Some of the material in this paragraph above feels like a slight repetition of the previous one. Can  they be combined/streamlined in some way?}

In this paper we study theoretically, within an entirely {\it ab initio} scheme, the processes undergone by phonons created by the interaction of a light DM particle in a polar semiconductor (target), until they reach the phonon collector. These results are essential to assess the feasibility of an innovative light DM detector and to evaluate (and then minimize) its non-ideal performances, i.e.\ the loss of phonon signal reaching the sensors. We focus on \al2o3\ which, as we discuss below, checks all the boxes of the ideal polar target for DM detection.  The phonon collector is taken to be Al, which, besides its suitable superconducting properties, presents high-quality interfaces with \al2o3. However, we emphasise that we focus exclusively on the phonon dynamics within the target, and that the properties of the collector only enter in the determination of the phonon transmission probability across the \al2o3/Al interface.

The structure of the paper is the following. In Section~\ref{sec:background} we give an overview of DM phonon scattering mechanisms and the detection scheme. In Section~\ref{sec:results} we present and discuss our results; specifically, (i)~we provide an exhaustive {\it ab initio} characterization of the phononic properties of the target/collector system chosen, i.e.~\al2o3/Al; (ii)~by time integration of the collision operator we discuss the validity of a number of approximations that can be adopted to describe the phonon downconversion and subsequent propagation. Finally, we present results from a newly developed energy-deviational Monte Carlo scheme, assessing the performances of the detector and discussing the feasibility of the detection strategy. In Section~\ref{sec:conc} we present our conclusions. The appendices provide a more insightful discussion regarding the employed methodologies.

\section{Phonon-based direct detection of light dark matter}
\label{sec:background}

The production of phonons offers an attractive channel for searching for light Dark Matter particles, especially for DM particles lighter than $\sim 1 \,\mathrm{MeV}$. The de Broglie wavelength of such light particles is much larger than the inter-atomic spacing in crystal targets, meaning that they can effectively couple with phonons in the material~\cite{Cox:2019cod,Trickle:2019nya,Trickle:2020oki}. 
While sub-MeV particles typically do not have sufficient kinetic energy to excite detectable nuclear recoils or electronic ionization, the production of phonons (with energies $\omega \gtrsim \mathrm{meV}$) is instead kinematically accessible. 

Depending on the model of DM, phonons may be produced by scattering or by the absorption of the DM particle directly by the target material. The rate of production of phonons with energy $\omega$ and momentum $\mathbf{q}$ can be obtained by convolving the DM-phonon interaction rate $\mathrm{d}\Gamma/\mathrm{d}\omega\mathrm{d}\mathbf{k}$ with the velocity distribution of DM in the Milky Way halo $f(\mathbf{v})$. Writing this rate per unit time and per unit target mass, we obtain~\cite{GriffinPRD18}:
\begin{equation}
\label{eq:DMRate}
\frac{\mathrm{d}R}{\mathrm{d}\omega\,\mathrm{d}\mathbf{q}} = \frac{1}{\rho_\mathrm{T}}\frac{\rho_\mathrm{DM}}{m_\chi} \int \mathrm{d}^3\mathbf{v} f(\mathbf{v}) \frac{\mathrm{d}\Gamma (\mathbf{v})}{\mathrm{d}\omega\,\mathrm{d}\mathbf{k}}\,.
\end{equation}
Here, $\rho_\mathrm{T}$ is the mass density of the target material. The local density of DM can be estimated from mass modeling of the Milky Way or from the dynamics of local stars~\cite{Read:2014qva,deSalas:2020hbh}, giving $\rho_\mathrm{DM} \approx 0.3 - 0.6 \,\mathrm{GeV}\,\mathrm{cm}^{-3}$. The DM velocity distribution in the Earth frame is typically assumed to take the form of a Maxwell-Boltzmann distribution (with typical velocity $v_0 \approx 220 \,\mathrm{km/s}$) and a cut-off at the Galactic escape velocity ($\sim 775\,\mathrm{km/s}$ in the Earth frame)~\cite{Baxter:2021pqo}. A number of public tools have been made available for the numerical calculation of the rates described by Eq.~\eqref{eq:DMRate}, including \texttt{PhonoDark}~\cite{PhonoDark} and \texttt{DarkELF}~\cite{DarkELF}.

As a concrete example, we will consider the case of a DM particle interacting via an ultra-light dark photon mediator. This effectively gives the DM particle a very small electromagnetic charge. Calculation of the DM-phonon scattering rate then proceeds in analogy with that of electron-phonon scattering, rescaled by the small effective charge of the DM~\cite{KnapenPLB18}. This scattering rate can also be related to the energy loss function (ELF) of the target material, which is well-studied both theoretically and experimentally~\cite{Knapen:2021run,Knapen:2021bwg}. One advantage of DM particles interacting via a dark photon is that they preferentially couple to optical phonons, which is particularly promising. For a fixed DM mass and velocity, optical phonons typically allow for a larger energy deposition compared to acoustic phonons, which comes purely from kinematic considerations and the fact that the optical phonons are gapped~\cite{Trickle:2019nya}. We therefore focus here on the production of a single optical phonon from DM interactions. The scattering of DM to produce two optical phonons is expected to be subdominant, and even kinematically inaccessible for sufficiently small DM masses~\cite{Campbell-Deem:2019hdx,Kahn:2020fef,Campbell-Deem:2022fqm}.

In order to detect the athermal phonon produced in the DM interaction, its energy must be transported to the surface of the target crystal and transferred to the phonon collector.
The overall process can be broken down in different stages (see the sketch in Figure~\ref{fig:detector}). At first the optical phonon generated by the DM particle must downconvert to a lower energy, long wavelength acoustic phonon. This is necessary for two reasons: (i)~high-energy optical phonons have typically very small group velocities and do not propagate efficiently; (ii)~if the phonon energy does not decrease, the elastic mismatch with the collector prevents its transmission. After downconversion, the resulting long wavelength phonon travels quasi-ballistically through the target (something made possible by the very low operation temperatures) until it reaches the collector. There, the phonon can be transmitted or reflected, with a probability that depends on the thermal boundary resistance between the target and the collector. If it is reflected, it will cross the whole length of the target twice (once backward, once forward after reflection from the backsurface) and attempt again to be transmitted. Typically, these multiple reflections occur when the initial downconversion process did not lower sufficiently the phonon energy, and the additional travelling through the target offers more opportunities for further downconversion events. The transmission probability at the interface is essentially ruled by mismatch of the elastic properties of the target and the collector, as at such low temperatures anharmonic effects can be likely neglected.

In the following, we study these phonon processes in more detail, beginning with a description of the phonon properties of \al2o3\ and Al, which we consider as promising materials for the target and phonon collector, respectively.

\section{Results and Discussion}
\label{sec:results}

\subsection{Materials: phonons and phonon-related properties for DM detection}
\label{sub:materials}

\noindent
{\bf Phonon dispersions.}~The starting point to approach any phonon-based detection scheme is the  phonon dispersion of the materials involved. Despite the increasing popularity of large-scale inorganic material databases, only a few of them include information about phonon properties (e.g. Materials Project~\cite{matproj}, almaBTE~\cite{almadb}, phonondb@kyoto-u~\cite{phononkyoto}). Here we compute the phonon dispersion of \al2o3\ and Al by finite differences in $4 \times 4 \times 4$ and $8 \times 8 \times 8$ supercells, respectively.
% \bjk{Do we also need to specify the q-mesh that we use? This is mentioned in the thermal conductivity section, but perhaps it should be here already.} \rr{No, we don't. There's no "mesh" as such in phonon calculations, only high-symmetry directions (indicated in the figure) and the number of points computed along each direction, but this just determines the graphical quality of the plot and has no real impact on the results.}

\begin{figure}[t]
\includegraphics[width=1.0\linewidth]{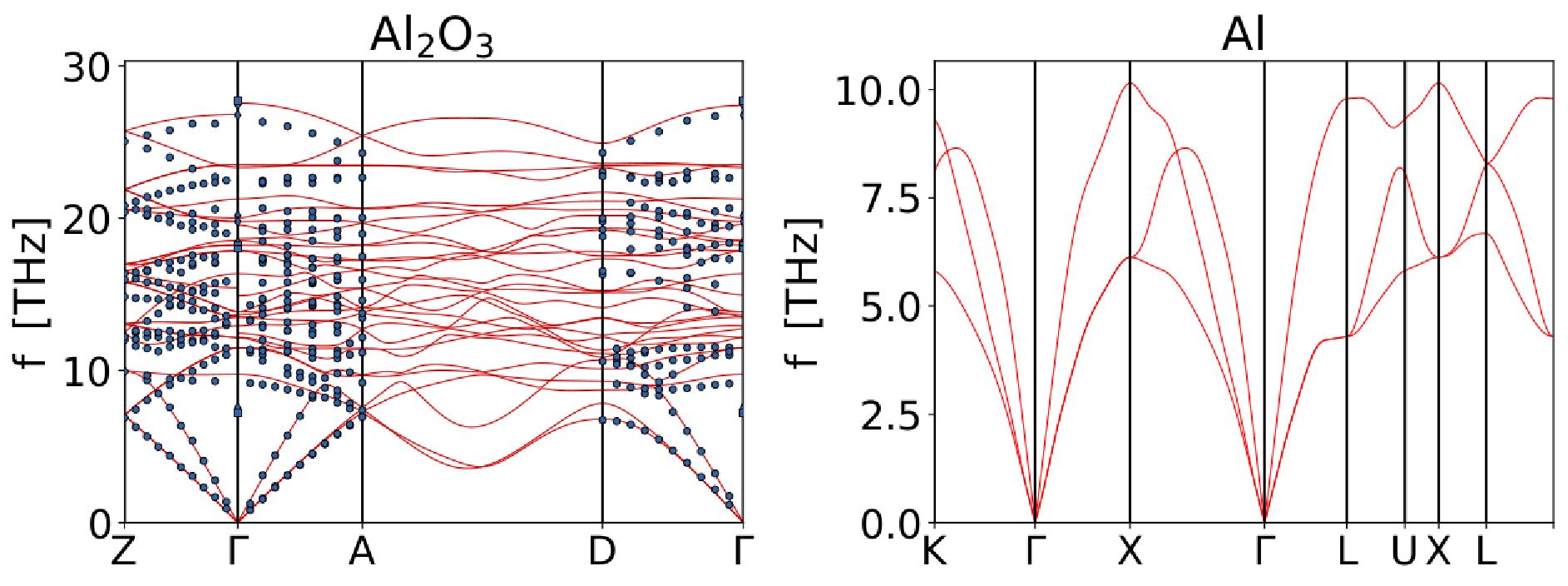}
       \caption{Phonon dispersions of \al2o3\ (left) and Al (right). The black dots indicate experimental measurement of Schober {\it et al.}~\cite{SchoberZPB93}. Note that $1\,\mathrm{THz}$ corresponds to an energy of around $4.14\,\mathrm{meV}$.}
\label{fig:bands}
\end{figure}

We use the \shell{phonopy} code~\cite{TogoPRB15} to generate all the inequivalent atomic displacements (in such supercells) that are required to compute the second order interatomic force constants (IFC) and then the dynamical matrix. The results for \al2o3\ and Al are shown in Figure~\ref{fig:bands}. A few things are worth noting: (i)~the bandwidth of the phonon dispersion of \al2o3\ covers the energy range that is expected to be important for sub-MeV DM detection, i.e.\ $\omega \lesssim 100 \,\mathrm{meV}$; (ii)~the target is anisotropic (e.g. the phonon bands along $\Gamma-Z$, $\Gamma-A$, and $\Gamma-D$ are different) and is thus suited to produce a daily modulation of the signal due to DM particles, which would provide further evidence of their detection; (iii)~the spectra of \al2o3\ and Al feature a considerable energy mismatch (phonons in Al cover barely one third of the bandwidth of \al2o3). This mismatch will play an important role when it comes to compute the  transmission probability of phonons when they hit the target/collector interface (see below).

\begin{figure}[t]
\includegraphics[width=1.0\linewidth]{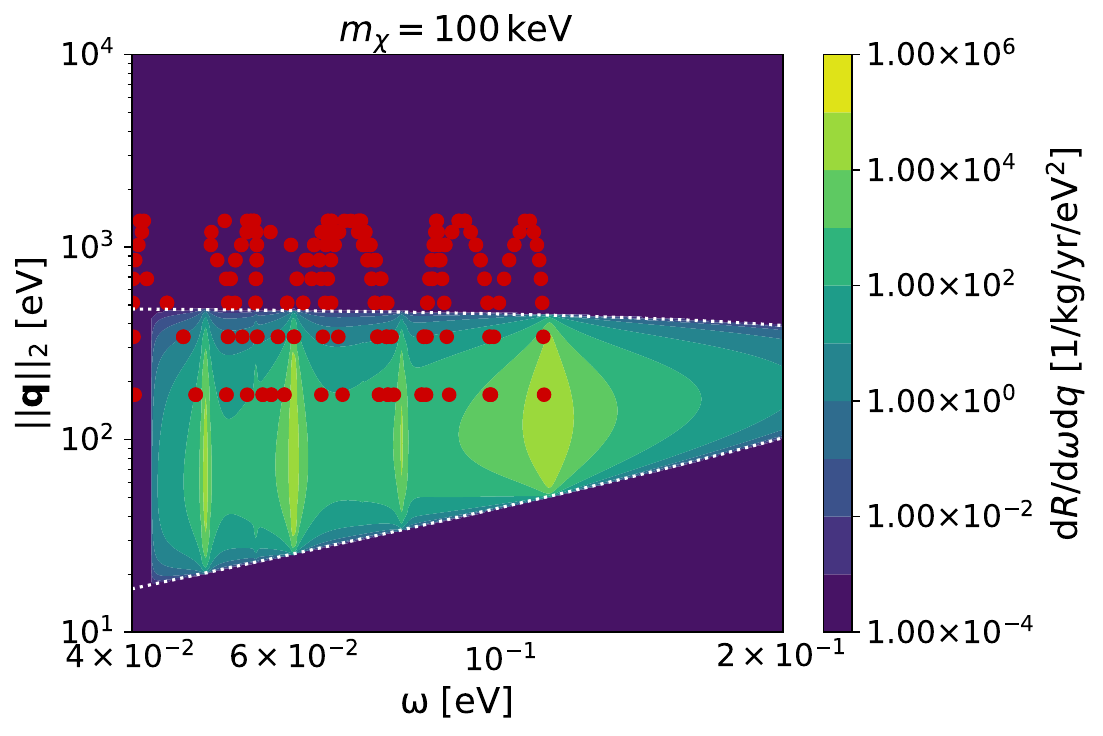}
       \caption{Spectral decomposition of DM phonon scattering rate per  
                unit of mass for \al2o3 in the $\langle 0001 \rangle$ direction for a DM particle of 100~keV. Here, we assume an ultra-light Dark Photon mediator and calculate the scattering rate using \texttt{DarkELF}~\cite{Knapen:2021bwg}. The dashed white lines mark the minimum and maximum momenta which are kinematically accessible for this DM mass. The overall normalisation of the rate is arbitrary here (but in general depends on the DM-scattering cross section). Red dots: available phonon states in the \al2o3\ target for a
                $\Gamma$-centered $17 \times 17 \times 17$ {\bf q}-mesh. %\bjk{Also - in DarkELF, they presumably use a different set of phonon dispersions compared to what we're using in the bulk of our paper. Should we therefore re-do the rate calculations using our dispersions for consistency? Just a thought.}
                %\rr{You are in principle right, but I0d say that citing DarkELF, like we do, is sufficient. Ab initio phonon dispersions are nowadays rather standard. Unless you deal with a very exotic material, dispersions of the samke material obtained with different code/approaches tend to qualitatively look like.}
               }
\label{fig:rates}
\end{figure}

It is important to highlight that having phonon states in the right
energy range for the detection of light and ultralight DM is only
a minimal requirement. In reality, one should convolve the DM-phonon scattering rate distribution of the DM particles incident on the target, as described in Eq.~\eqref{eq:DMRate}. 
%With this, it is possible to determine the rate of phonon production for states available in the target. 
As a concrete example, we show in Figure~\ref{fig:rates}, the expected rate of production $\mathrm{d}R/\mathrm{d}\omega\mathrm{d}k$ of phonons with energy $\omega$ and momentum $k$, due to DM-phonon scattering. We perform the calculation using the \shell{DarkELF} code~\cite{Knapen:2021bwg}, assuming a fermionic DM particle of mass $m_\chi = 100 \,\mathrm{keV}$ interacting via an ultralight dark photon mediator. For simplicity, we assume phonon production only along the ordinary direction (i.e.\ the $\langle 0001 \rangle$ direction) of \al2o3.

At a given phonon energy, the rate of phonon production is non-zero only for a finite range of momenta. This range is set by kinematics; requiring conservation of energy and momentum in the scattering process leads to the constraint $\omega \leq \lVert \mathbf{q}\rVert v_\chi - \lVert \mathbf{q}\rVert^2/2m_\chi$, where $v_\chi$ is the incoming DM velocity. The maximum value of $v_\chi$ is set by the sum of the Earth's velocity and the Milky Way escape velocity ($v_\mathrm{max} \approx 775\,\mathrm{km/s}$)~\cite{Green:2017odb,Baxter:2021pqo}, and this in turn sets the maximum momentum for phonon states which can be produced in the scattering event. Overplotted as red dots in Figure~\ref{fig:rates} are the available phonon states in the $\mathbf{q}$-mesh we use in this work, demonstrating that some of these are accessible.

%In reality, one should map the energy and momentum
%of DM particles and make sure that such states are available in the target.

\smallskip
\noindent
{\bf Thermal conductivity.}~The detection process begins with a DM particle scattering in the target to produce an optical phonon. For the detection to be efficient, we need this phonon first to downconvert to longer wavelength, quasi-ballistic phonons, which, in turn, should travel long distances  with minimal scattering, so that a large signal reaches the target/collector interface. These two requirements are to some extent contradictory: downconversion relies on a certain level of anharmonicity, but anharmonicity is also the main cause for phonon scattering. To have a first rough idea, we computed the thermal conductivity of \al2o3\ by solving the Peierls-Boltzmann Transport Equation (PBTE)~\cite{ZimanEP,LandonJAP14,FugalloPS18}, using as inputs the harmonic and anharmonic interatomic force constants (IFCs) obtained from DFT calculations.
The third-order anharmonic IFCs were calculated in the same supercells used for the harmonic ones, considering interactions up to 7$^\text{th}$ neighbors. Scattering from isotopes is accounted for within the model of Tamura~\cite{TamuraPRB83}, considering the natural isotopic population of O [Al has only one stable isotope, $^{27}$Al, while O has three: $^{16}$O (99.8\%), $^{17}$O  (0.038\%), and $^{18}$O (0.205\%)]. We solve the PBTE on a $17 \times 17 \times 17$ {\bf q}-point grid with the almaBTE code~\cite{CarreteCPC17} (see \hyperref[sec:Methods]{Methods} for full details of the calculations).

\begin{figure}[t]
\includegraphics[width=1.0\linewidth]{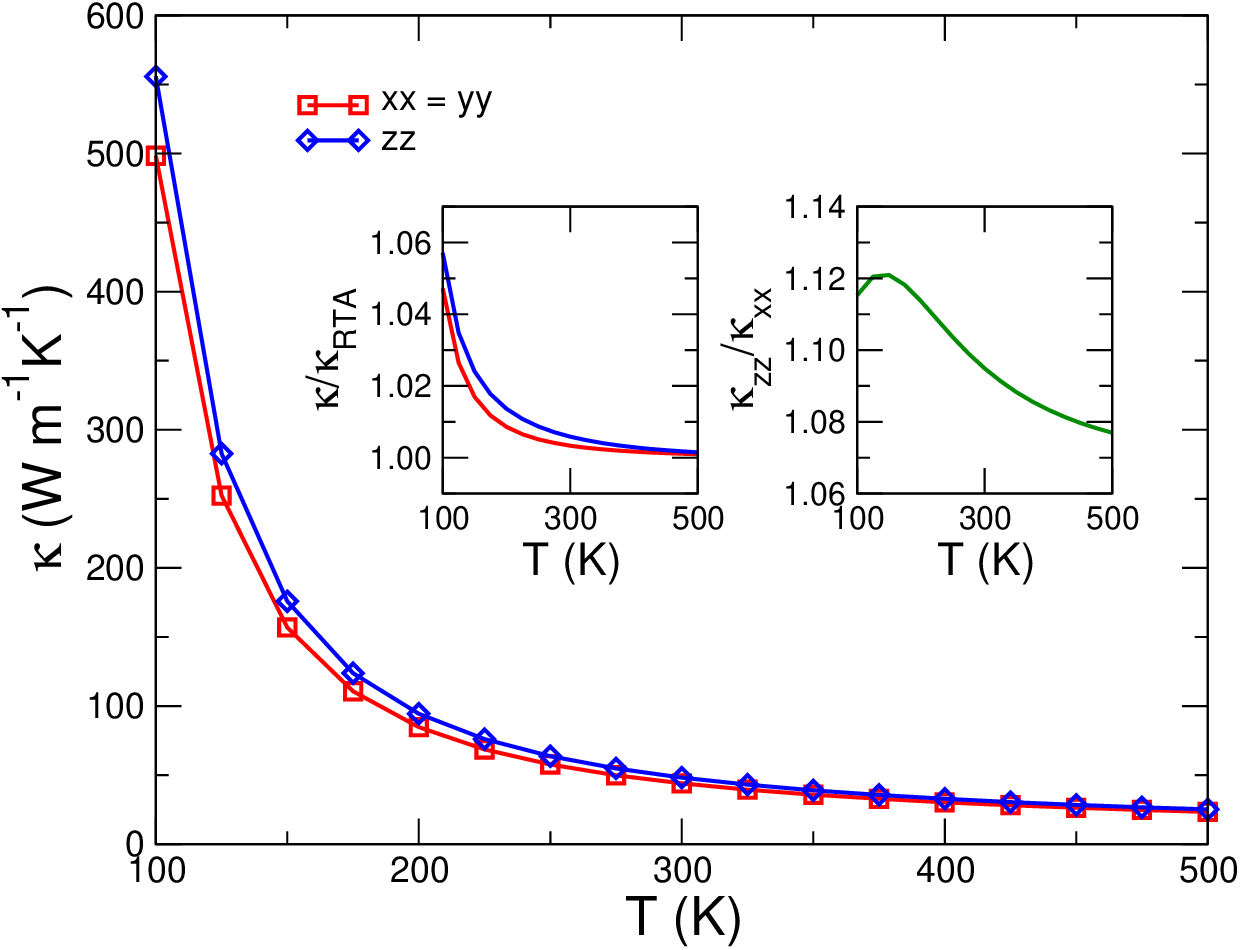}
       \caption{Thermal conductivity of \al2o3 as a function of temperature.
                Insets: ratio between the full solution of the PBTE and the one obtained within the RTA (left); anisotropy, defined as $\kappa_{zz}/\kappa_{xx}$ (right).
               }
\label{fig:kappa}
\end{figure}

We obtain a thermal conductivity at 300~K of $\kappa_{xx}=\kappa_{yy}=44.0$ and $\kappa_{zz}=48.2$~W~m$^{-1}$~K$^{-1}$, which increases up to  500-550~W~m$^{-1}$~K$^{-1}$ when the temperature decreases down to 100~K, as phonon-phonon scattering processes become much weaker (see Figure~\ref{fig:kappa}). These results agree well with those previously obtained by Dongre and coworkers~\cite{DongreMRSComm18}. Clearly, at the temperatures relevant for the present detection scheme, $T \approx 10$~mK, the thermal conductivity will be much higher. However, a reliable estimate of ${\bm \kappa}$ at such low temperatures is beyond current computational capabilities, because of the extraordinarily fine {\bf q}-point mesh necessary to properly capture very long wavelength phonons, which become increasingly important as the temperature decreases.

These results simply provide a general indication that \al2o3\ has a medium room temperature thermal conductivity for a single-crystalline solid. What matters for the detection scheme under scrutiny is how a specific optical phonon downconverts and how efficiently the (ideally quasi-ballistic) phonons resulting from the downconversion process propagate. These questions will be specifically addressed below.

The insets of Figure~\ref{fig:kappa} report two results that are 
important for the forthcoming discussion. On the left hand-side we plot
the ratio between the values coming from a full solution of the PBTE
and those obtained within the simpler Relaxation Time Approximation (RTA)~\cite{LundstromBOOK2002} (see Sec.~\ref{sub:timeint}).
Corrections to the RTA made by the full solution of the PBTE are essentially
negligible at room temperature, but they increase up to $\approx 6$\% at
100~K, a temperature which is still four orders of magnitude larger than
the operating temperature of a phonon-based DM detector. The inset on the right
displays the thermal conductivity anisotropy, $\kappa_{zz}/\kappa_{xx}$,
which, as discussed above, is important when it comes to the daily modulation
of the detected signal. 
\\

% say something about q-point convergence
%
% can it be that we cannot be converged for kappa at cryogenic T, but
% the Harrelson stuff is okay? Ask Marti

\begin{figure}[t]
\includegraphics[width=1.0\linewidth]{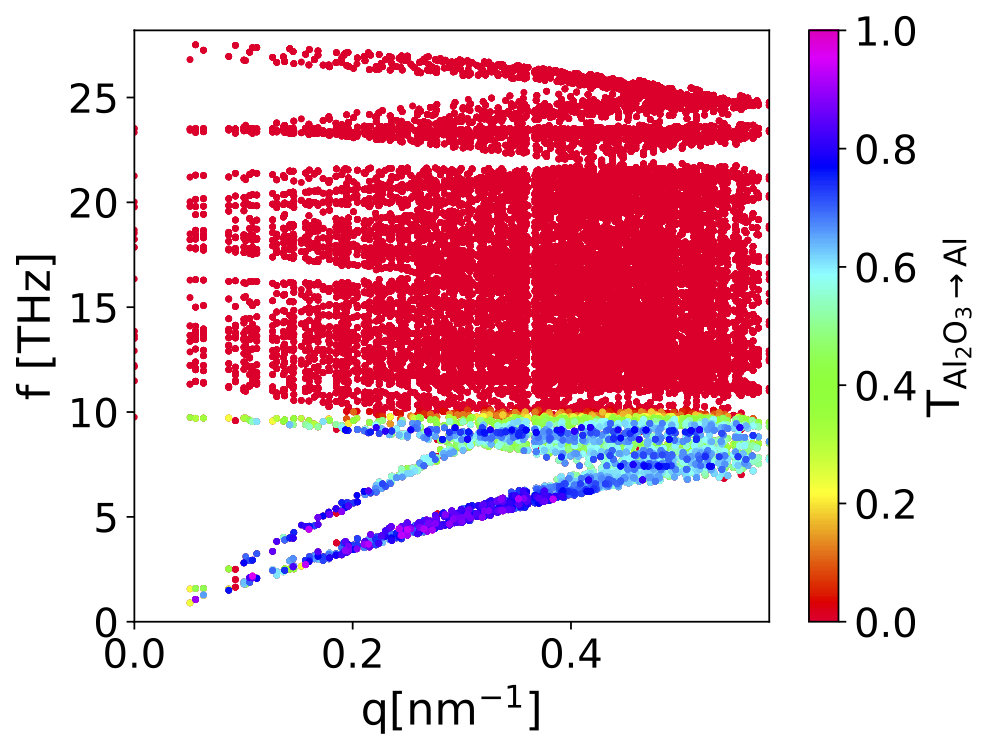}
       \caption{Phonon dispersion of \al2o3\, where a color code indicates the value of the transmission probability computed within the DMM for a $\langle0001\rangle$ Al\textsubscript{2}O\textsubscript{3}-$\langle111\rangle$ Al interface.
       }
\label{fig:dmm}
\end{figure}

\noindent
{\bf Thermal boundary resistance.}~When the heat flux hits the target/collector interface, part of it will be transmitted with a certain probability, $\alpha$, and part will be reflected. The fact that not all phonons are transmitted can be understood in terms of the mismatch of the elastic properties of the materials on the two sides of the interface. This is easy to understand looking at the phonon dispersions of \al2o3\ and Al displayed in Figure~\ref{fig:bands}: it is clear that phonons of the target (\al2o3) with $f > 10$~THz cannot be transmitted to the collector (Al), because there are no available phonon states at that energy. The situation is further complicated when considering that (i)~even when a finite vibrational density of states exists on the two sides of the interface, the transmission probability depends on the degree of coupling of the states involved and is in general lower than unity, and (ii)~the effect of interface anharmonic scattering can open additional transmission channels~\cite{CarreteNanoscale19}. Indeed the development of a microscopic theory of thermal boundary resistance (TBR) is one of the most active areas in thermal science~\cite{SwartzRMP89,ChenRMP22}.

In the following, we will make use of one of the most common phenomenological models for the calculation of the TBR, the so-called  diffuse mismatch model (DMM) ~\cite{SwartzRMP89} in a full-band formalism~\cite{ReddyAPL05,LarroqueJAP18}. Within DMM, all phonons undergo fully diffusive scattering at the interface\footnote{That is, the perpendicular momentum of the outgoing phonon is fully randomized under the energy conservation constraint. This is in contrast with specular scattering in which the outgoing momentum is uniquely determined by the input phonon.} and they lose memory of their origin after being scattered. In Figure~\ref{fig:dmm} we plot the transmission probability across an \al2o3/Al interface for each phonon on the \al2o3\ side. As anticipated, the transmission of phonons with $f > 10$~THz is suppressed, simply because no states at that energy are available on the Al side. Transmission at lower frequencies is larger, though a finite reflection probability exists in most cases.
% Previously, $\alpha$ has been taken to be 0.5, regardless of the phonon mode involved.

\subsection{Time integration of the collision operator}
\label{sub:timeint}

{\bf The Peierls-Boltzmann Transport Equation.}~As the common choice for target materials are high-quality single-crystal polar semiconductors, the phonon dynamics can be in principle decoupled from the electronic one, and thus can be described through the deviational Peierls-Boltzmann Transport Equation (PBTE)~\cite{ZimanEP,LandonJAP14}: 
\begin{equation}
	\label{PBTE}
	\frac{\partial n^d_i}{\partial t} + \mathbf{v}_i \cdot {\nabla_r} n^d_i + \frac{\partial n^0_i}{\partial T} \mathbf{v}_i \cdot {\nabla_r} T_{\mathrm{ref}}=  \left.\frac{\partial n_i}{\partial t} \right|_\mathrm{collision}\,.
\end{equation}
This approach is {\it deviational} in the sense that rather than describing the evolution of the total phonon population in the target, $n$, which normally results in poor signal-to-noise ratios, it deals with the deviation of the phonon population from equilibrium, $n_d = n - n_0$. The equilibrium distribution $n^0$ is well described by a Bose-Einstein distribution at the target temperature $T_{\mathrm{ref}}$, while the deviation $n^d$ is the DM-generated (athermal) phonon contribution.
%^Here, the total phonon distribution in the target ($n$) has been decomposed into the equilibrium distribution ($n^0$, which is well described by a Bose-Einstein distribution at the target temperature, $T_{\mathrm{ref}}$) and the DM generated (athermal) phonon contribution (i.e. $n^d$)
Here, $\mathbf{v}_i$ is the group velocity of the $i$-th mode; and $\left.\frac{\partial n_i}{\partial t} \right|_\mathrm{collision}$ is the collision operator, modeling the change in the phonon distribution due to all the possible phonon scattering mechanisms, e.g.\ phonon-phonon processes, isotopic and boundary scattering. 

Due to its complexity, the full nonlinear phonon collision operator 
presents significant challenges when solving the PBTE, making its 
practical use limited in most situations. A common approach 
consists in reducing it to a more tractable form by linearizing it,
i.e.\ by considering only small deviations from equilibrium, $n^0 \gg n^d$.
The  collision operator reduces then to:
\begin{equation}
        \label{eq:collisionAnd}
        \frac{\partial n_i^d}{\partial t}\Big|_{\mathrm{collision}} = A_{ij} n_j^d
\end{equation}
where $A_{ij}$ is the linearized collision operator.

Sometimes, the PBTE is further simplified using the so-called 
relaxation time approximation (RTA), where it is assumed that each 
phonon mode relaxes to equilibrium independently with a characteristic time, $\tau_{i}$. This translates to 
supposing that only the involved mode, $i$, is slightly out of equilibrium,
while the other modes remain in equilibrium (i.e. $n_j = n_j^0 \; \forall\; j\neq i$). 
In this way, the RTA collision operator, $A_{ij}^\mathrm{RTA}$, can be expressed in terms of the full linearized collision operator as:
\begin{equation}
    A_{ij}^\mathrm{RTA} = A_{ij}\delta_{ij} = -\frac{n^d_i}{\tau_i}\delta_{ij},
\end{equation}
where the last is the most common expression for the RTA.

Despite its simplicity, which has earned it a great renown in 
the field of phonon and lattice thermal calculations (particularly 
when fully {\it ab initio} calculations of the thermal conductivity
first became possible~\cite{WardPRB09}) the RTA has several limitations.
It is easy to see that, under this approximation, the collision operator 
does not conserve heat flux to any degree, deeming all the three-phonon 
processes resistive. Such a flaw causes the RTA to fail for cases in 
which non-resistive processes, i.e.\ those that mostly
redistribute the phonon population, play a non-negligible role. 
This is the case of e.g.\ diamond and 2D materials, 
for which the RTA has been observed to provide quite a poor description 
of thermal properties, even at room temperature. However, at sufficiently 
low temperatures this situation becomes common to all materials and predictions
based on the RTA are in general unreliable~\cite{WardPRB09,CepellottiNatComm15,LindsayJAP19}.

The target semiconductor material used here, \al2o3, is a good example: the error induced by the RTA, when compared to a full solution of the PBTE, is
negligible at room temperature, but it already increases to $\approx 6$\% at $T=100$~K (see the discussion above). Although we provide a careful validation of the RTA below, these considerations by themselves already suggest that the RTA is most likely bound to yield an unreliable description of a  phonon-based detection scheme that is expected to operate at cryogenic temperatures.

There is another limitation of a pure RTA operator which is specific
of this kind of setup that is important to discuss. The central
assumption of the RTA is that each nonequilibrium (athermal) phonon 
mode returns to equilibrium independently from the rest. It follows
from this approximation that redistribution among the different modes
is not permitted and thus no flux of phonons will be generated if the 
initial state is fully located at a standing mode (i.e.\ 
$\mathbf{v}_\lambda = \mathbf{0}$). A more realistic approach 
in our case would be to allow a relaxation of the initial mode 
$K$ to other modes, which then thermalize (i.e.\ allowing only 
initial state out-coupling). This generalization of the RTA, which 
we dub pseudo-RTA, serves in principle our purposes and can be 
expressed in terms of the linearized collision operator as: 
%\bjk{What do we mean here by the square brackets?}\mrm{It is an Iverson bracket, it is somehow a fancy kronecker delta to write complex conditions.}
\begin{equation}
    \label{eq:pseudoRTA}
	A_{ij}^{\mathrm{pseudo-RTA}} = A_{ij}  [i = j ~\mathrm{or}~j = K].
\end{equation}

The linearization of the PBTE is a much more widespread approximation
than the RTA and virtually all the codes that calculate the thermal
conductivity from {\it ab initio} computed IFCs rely on it,
even when beyond-RTA solutions are pursued~\cite{LiCPC14, TogoPRB15, CarreteCPC17}. The validity of this approximation, which is not normally the source of major problems, should nonetheless be addressed in detail in the case we are studying: it is true that we assume that only one phonon mode is excited from the detection of DM, but at the same time is driven to a strongly out-of-equilibrium condition. It is difficult to say which of these two factors dominates and thus the reliability of a linearized solution, as well as of the pseudo-RTA, are carefully assessed below.
\\

\noindent
{\bf Different treatments of the collision operator.}~As a first approach to the problem of phonon-based DM detection, we present results of the time integration of the collision operator in our \al2o3/Al system. These
calculations serve to discuss the validity of the two approximations
discussed above, i.e. the linearization and the RTA, which
have been previously used in the literature~\cite{HarrelsonArXiv21}.

Our starting point is a simplistic expression of the heat flux transmitted from the target material to the collector~\cite{SwartzRMP89}:
% \begin{multline}
%        \label{eq:GrossFlux}
        % J _{\mathrm{gross}}(t) =\frac{1}{N_{\mathrm{q}}V_{\mathrm{uc}}}\sum_i  \alpha_i J_{\mathrm{proj},i} (t) = \\ \frac{1}{N_{\mathrm{q}}V_{\mathrm{uc}}}\sum_i \alpha_i \hbar\omega_i n^d_i(t) \mathbf{v}_i\cdot \mathbf{u} ~ [\mathbf{v}_i\cdot \mathbf{u} > 0],
% \end{multline}
\begin{eqnarray}
        \label{eq:GrossFlux}
        J _{\mathrm{gross}}(t) & = & \frac{1}{N_{\mathrm{q}}V_{\mathrm{uc}}}\sum_i  \alpha_i J_{\mathrm{proj},i} (t) \nonumber \\ &=& \frac{1}{N_{\mathrm{q}}V_{\mathrm{uc}}}\sum_i \alpha_i \hbar\omega_i n^d_i(t) \mathbf{v}_i\cdot \mathbf{u} ~ [\mathbf{v}_i\cdot \mathbf{u} > 0],
\end{eqnarray}
where $N_{\mathrm{q}}$ is the number of  {\bf q}-points, $V_{\mathrm{uc}}$ is the volume of the unit cell, $\hbar$ is the reduced Planck constant, $\omega_i$ is the frequency of the $i$-th phonon mode, $\alpha_i$ is the transmission coefficient of the given mode to the collector material, $\mathbf{v}_i$ is the group velocity, and $\mathbf{u}$ is a normalized vector pointing out of the target into the collector; $n^d(t)$ is the deviational phonon distribution, i.e. the difference between the nonequilibrium and the equilibrium distribution function, $n - n^0$.

\begin{figure}[t]
\includegraphics[width=1.0\linewidth]{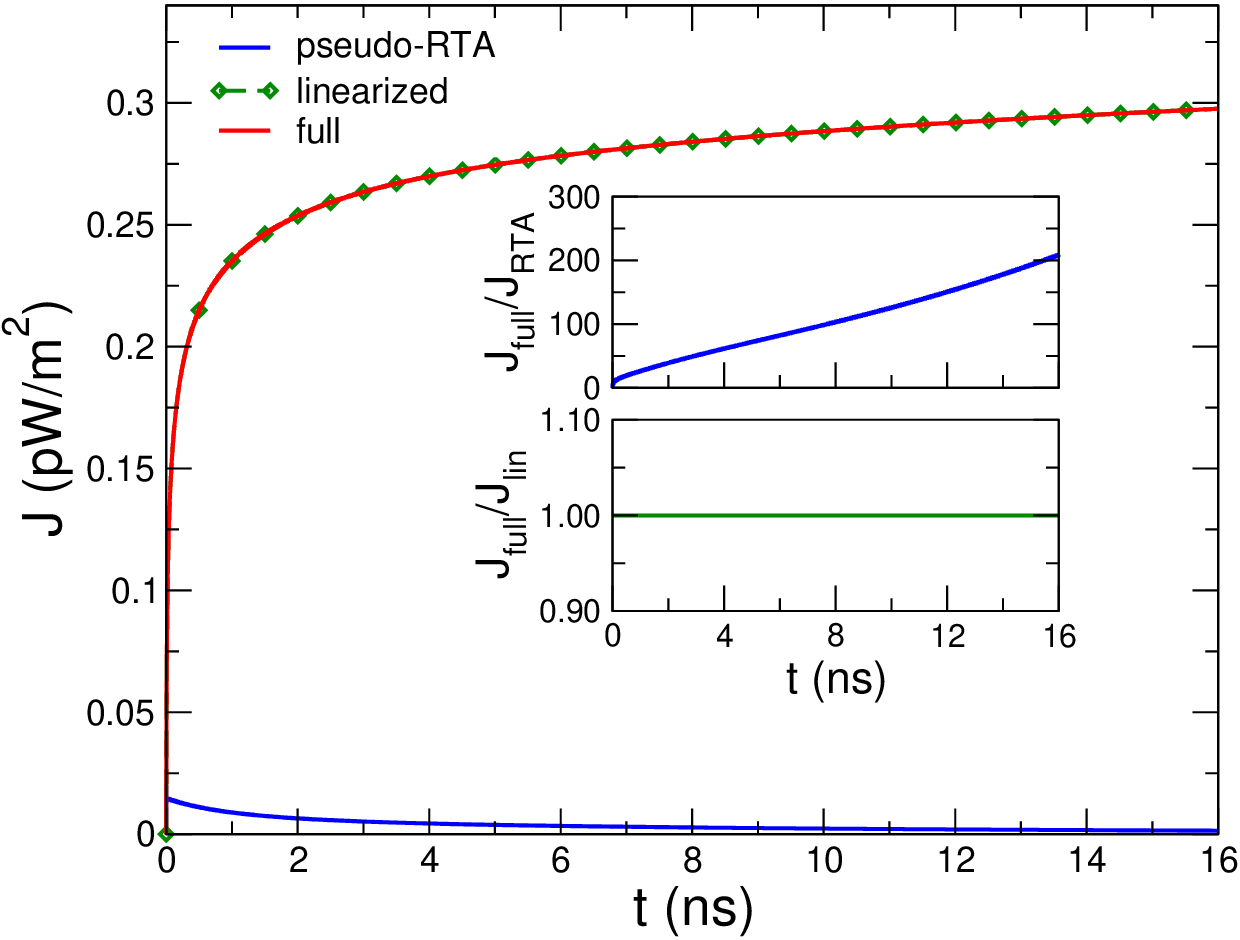}
       \caption{Time evolution of the heat flux calculated with different approaches to the collision operator. The insets show the relative error in the heat flux associated with the use of the pseudo-RTA or the linearization of the collision operator.}
\label{fig:rta-lin-full}
\end{figure}

In Figure~\ref{fig:rta-lin-full} we plot the time evolution of the 
projected heat flux, comparing the different approaches used to treat 
the collision operator. As it can be seen, relying on the pseudo-RTA
results in a severe underestimation of the heat flux.
This is not surprising because, as discussed above, the RTA considers all
phonon collisions to be resistive, while at the very low temperatures 
here considered the majority of phonon-phonon process simply redistribute 
momentum, roughly conserving the heat flux. The two beyond-RTA solutions,
on the other hand, describe correctly the phonon scattering dynamics and
yield larger fluxes, thus increasing the probability that the signal
originated from the DM interaction reaches the collector and can be successfully
detected.

Remarkably, we find virtually no difference between the solution obtained from the full and the linearized collision operator (lower inset of Figure~\ref{fig:rta-lin-full}). This is a very important conclusion, not only because the linearized solution is computationally less expensive, but also because this opens up the possibility to use Monte Carlo algorithms (which are considered the gold standard in solving the BTE, but are only applicable when linearization holds). Notice, incidentally, that the results in Figure~\ref{fig:rta-lin-full} were obtained assuming a volume of 1000~nm$^3$, which is a very small value compared to any realistic target sample. Therefore, these conclusions are particularly robust.

Notice that in the beyond-RTA solutions we obtain a build up of the heat flux, i.e.\ $J$ increases with time, due to the subsequent energy downconversion of phonons, which leads to the increase in the population of lower frequency modes with higher group velocities. Such a build up is absent in the pseudo-RTA, where phonons are allowed to downconvert only once.
% See the Supplementary Video~1 for the flux build up.

As linearization seems to be a valid approximation, at least in the limit of a single-phonon perturbation signal, this is the framework assumed from here on. Likewise, any RTA-based approach must be discarded.

\subsection{Energy-deviational Monte Carlo simulations}
\label{sub:mc}

The results presented in Section~\ref{sub:timeint} are important because they already capture important physical effects that are relevant for phonon-based detection of DM. Above all, however, they permitted us to assess different approaches to the collision operator, leading us to conclude that it can be safely linearized, while any RTA-derived description at such low temperatures is necessarily inaccurate. 

However, the main limitation of the theoretical framework adopted thus far is that the spatial dimension is completely neglected. Therefore, not even the simplest device geometries can be described. In this scheme, a DM particle is detected and the time evolution of the phonons generated in the target can be studied, but the value of the heat flux is the same at any point in space (or, equivalently, the propagation speed is infinite). In this way, not only are nuances of phonon propagation, e.g.\ multiple reflections, completely unaccounted for, but even a basic question, such as what the optimal target thickness is, remains unanswered.

\begin{figure}[t]
\includegraphics[width=1.0\linewidth]{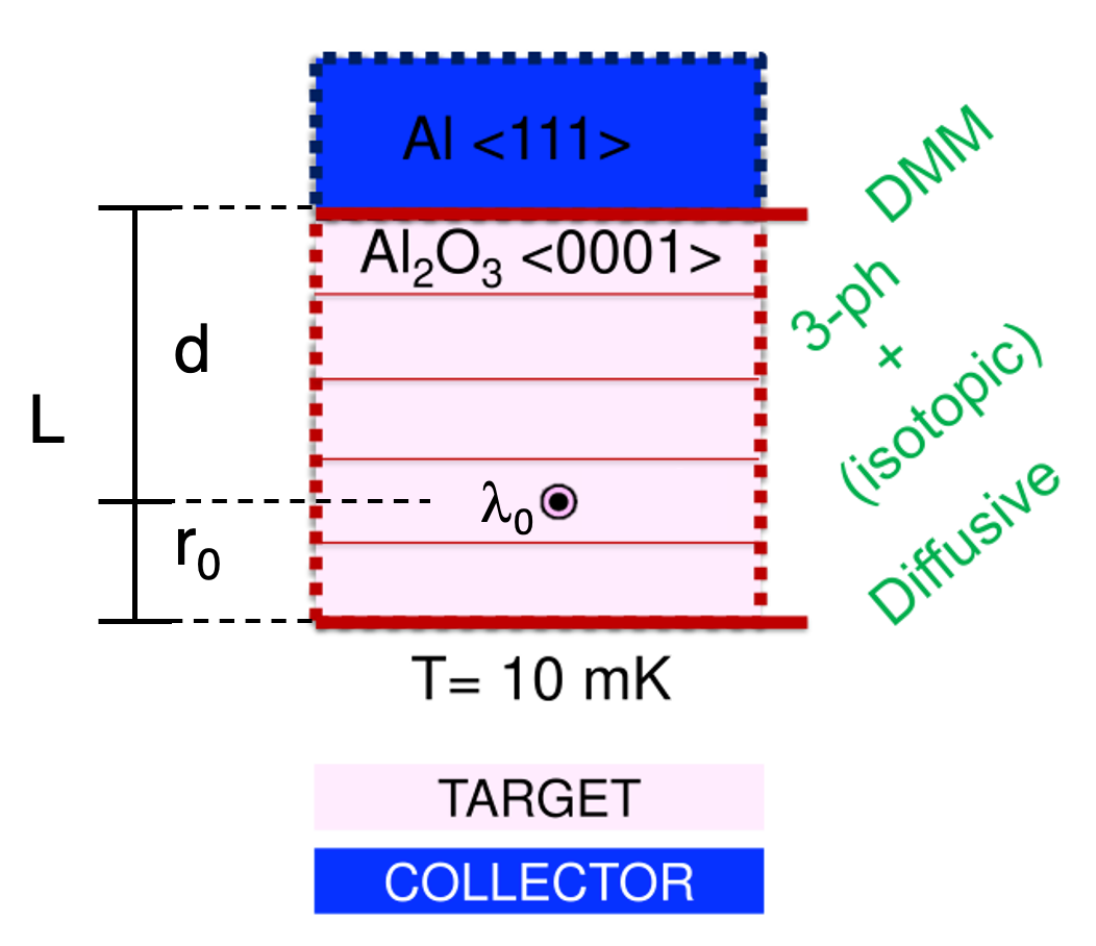}
       \caption{Sketch of the geometry used in the MC simulations. The computational system is composed of the target (pale rose) at \SI{10}{\milli\kelvin}, made of $\langle0001\rangle$ Al\textsubscript{2}O\textsubscript{3}, divided in several computational boxes along transport direction---note here that lateral boundaries and their effects are disregarded---, with a perfectly diffusive boundary, i.e. modeled with DMM, in one side ($x=\SI{0}{\nano\meter}$) and a perfectly diffusive interface with $\langle111\rangle$-Al (navy) at the other side ($x = L\;\SI{}{\nano\meter}$). The initial phonon is generated with a chosen mode $\lambda_0$ at $x=r_0$, or what is the same $d$\;\SI{}{\nano\meter} away from the interface. Phonons within the target suffer anharmonic phonon scattering processes, as well as isotopic one, if selected.}
\label{fig:mc_sketch}
\end{figure}

To circumvent this limitation, we developed a 3D deviational-energy Monte Carlo method that relies on a description of the linearized collision operator beyond the RTA (see Methods), whose accuracy has been validated in the previous section.
% The approach that we follow is dubbed {\it deviational-energy} because a direct implementation in terms of the phonon population:
%\begin{equation}
%    \frac{\partial n_i}{\partial t} + {\bm v}_i \cdot \nabla n_i = \left.\frac{\partial n_i}{\partial t}\right|_\mathrm{collision}
%    \label{eq:mc}
%\end{equation}
%would suffer of two shortcomings: (i)~a low signal-to-noise ratio and (ii)~energy conservation would require extra algorithms that can bias the solution.
In this framework, and after linearizing the collision operator, Eq.~\ref{PBTE} becomes:
\begin{equation}
    \frac{\partial n^d_i}{\partial t} + \mathbf{v}_i \nabla n^d_i + \frac{\partial n^0_i}{\partial T} \mathbf{v}_i \cdot \nabla T_\mathrm{ref} = \sum_j{A_{ij}n^d_j}\, .
    \label{eq:mc-dev}
\end{equation}
Also, we make the transformations $f_i = \hbar \omega_i n_i$ and $B_{ij} = \frac{\omega_i}{\omega_j} A_{ij}$ and obtain the final deviational-energy formalism:
\begin{equation}
    \frac{\partial f^d_i}{\partial t} + \mathbf{v}_i \nabla f^d_i + \frac{\partial f^0_i}{\partial T} \mathbf{v}_i \cdot \nabla T_\mathrm{ref} = \sum_j{B_{ij}f^d_j} \, ,
    \label{eq:mc-dev-ene}
\end{equation}
which naturally guarantees energy conservation, without the need for additional algorithms that might bias the solution.

The device setup that we study is illustrated in Figure~\ref{fig:mc_sketch}. The first question that we address
is to what extent the energy and momentum of the optical 
phonon created as a result of the DM scattering event matters. To this 
end we selected phonons of \al2o3\ that, according to the maps of 
Figure~\ref{fig:rates}, can be excited by DM and we studied their 
evolution.
Specifically, we have selected the following highly interacting phonon modes: high-energy-low-momentum ($E=\SI{0.11}{\electronvolt},~\lVert\mathbf{q}\rVert=\SI{171}{\electronvolt}$), high-energy-high-momentum ($E=\SI{0.11}{\electronvolt},~\lVert\mathbf{q}\rVert=\SI{342}{\electronvolt}$), low-energy-low-momentum ($E=\SI{0.06}{\electronvolt},~\lVert\mathbf{q}\rVert=\SI{171}{\electronvolt}$), and high-energy-high-momentum ($E=\SI{0.06}{\electronvolt},~\lVert\mathbf{q}\rVert=\SI{342}{\electronvolt}$) modes, all of them parallel to the $\langle0001\rangle$ crystalline direction.

\begin{figure}[t]
\includegraphics[width=1.0\linewidth]{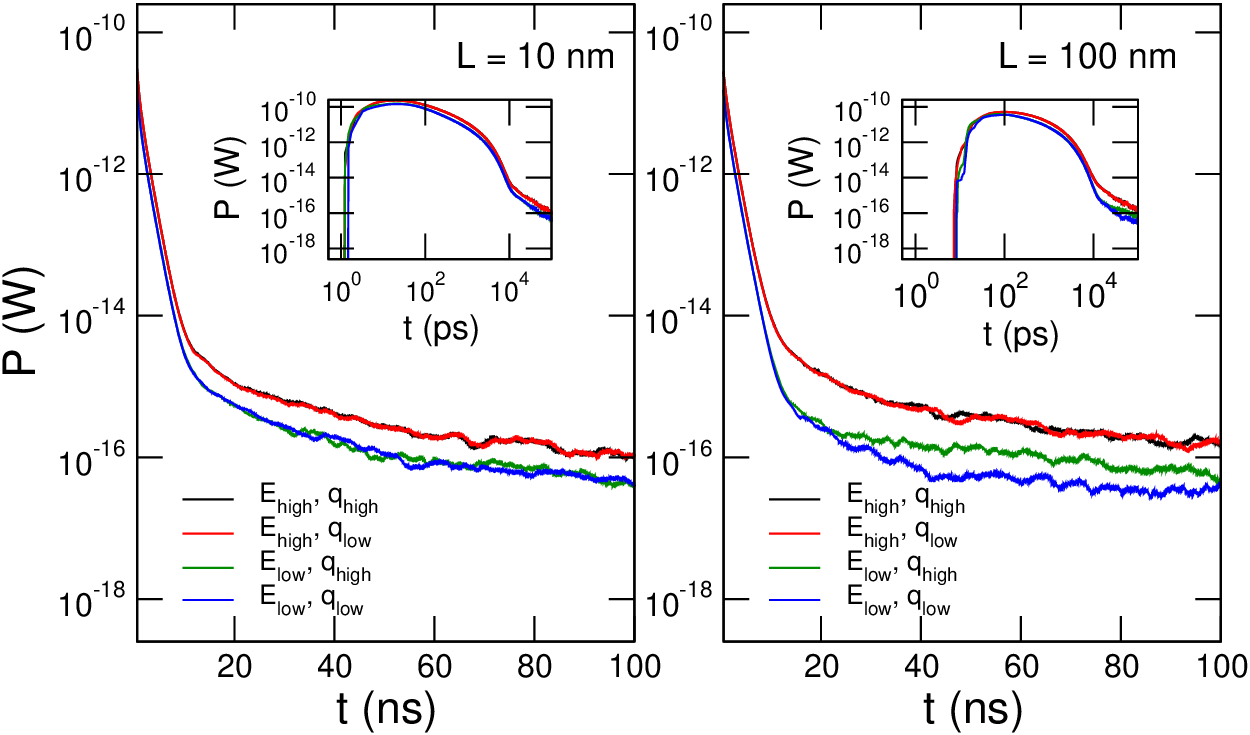}
       \caption{MC-computed power at the collector side (Al) of the interface as function of time, for the geometry shown in Fig.~\ref{fig:mc_sketch} with a length of \SI{10}{\nano\meter} (left) and \SI{100}{\nano\meter} (right). We compare the effect on the signal of the initial phonon state; namely, all possible combinations, of low ($E= \SI{0.06}{\electronvolt}$) and high ($E=\SI{0.11}{\electronvolt}$) energy states, and low ($\lVert\mathbf{q}\rVert=\SI{171}{\electronvolt}$) and high ($\lVert\mathbf{q}\rVert=\SI{342}{\electronvolt}$) quasi-momentum states parallel to the \al2o3-$\langle0001\rangle$ direction. Insets: log-log plots highlighting the peak of the power at very short values of $t$.
       All the simulations displayed in this panel include isotopic scattering; the initial phonon is in the same relative position ($r_0/L$) and $t=0$ corresponds to the instant of DM scattering and phonon creation.}
\label{fig:J_Eq}
\end{figure}

The results are plotted in Figure~\ref{fig:J_Eq} for two different device sizes, $L$, and indicate that higher energies provide higher signal, i.e.\ larger heat fluxes and thus power. Conversely, phonon momentum seems not to be important at all. Indeed, we considered the time evolution of two phonon states of essentially the same energy, $\omega \sim 0.11$~eV, but different {\bf q}-vectors, finding that the differences in the generated power are negligible ($L = 10$~nm, black and red curves in the left panel of Figure~\ref{fig:J_Eq}) or small ($L = 100$~nm, green and blue curves in the right panel). Notice that we focus our analysis on the tail of the signals, considering the typical response time of TES. Signal peaks, which are not significative for our purposes due to the aforementioned reasons, can be appreciated in the log-log plots in the insets.

%\bjk{How does the size of the device affect the signal? I don't think we've specifically addressed this in the text yet.}\mrm{What do you mean by size? If you refer to L, I think it is well discussed bellow, if you refer to actual 3D size, it is tricky as it is implemented only one dimension is really accounted, while the rest are supposed to be big in comparison.}.

\begin{figure}[t]
\includegraphics[width=1.0\linewidth]{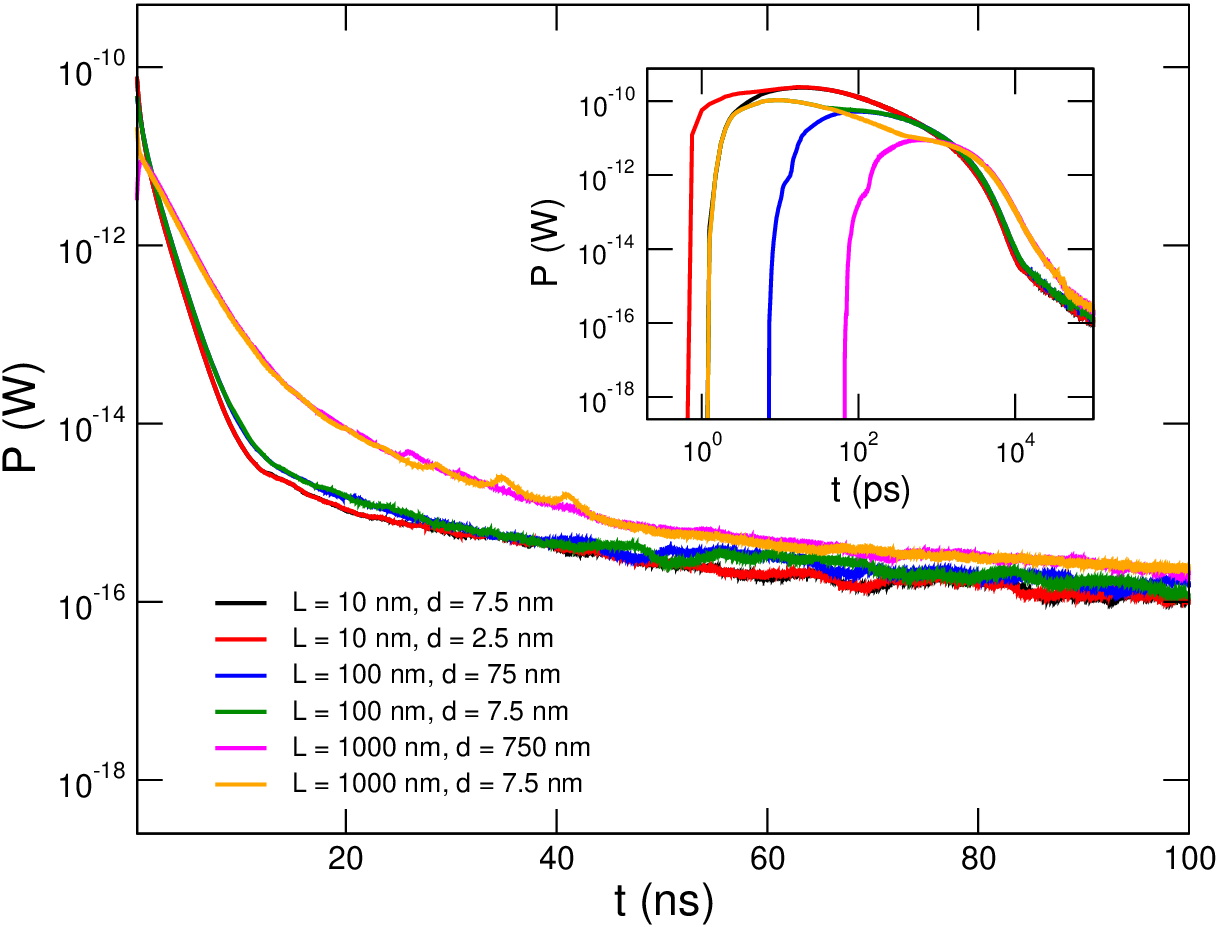}
       \caption{MC-computed power at the collector side (Al) of the interface as function of time, for a Fig.~\ref{fig:mc_sketch}'s geometry, comparing the effect on the collected signal of the device length (L=10,100,\SI{1000}{\nano\meter}), and the position (r\textsubscript{0}) in which the single particle is generated. The initial state for all cases is one of high energy (f=\SI{0.11}{\electronvolt}) and low momentum ($\lVert\mathbf{q}\rVert$=\SI{171}{\electronvolt}) parallel to the \al2o3-$\langle0001\rangle$ direction. Inset: log-log plot highlighting the peak of the power at very short values of $t$. All the simulations displayed in this panel include isotopic scattering; $t=0$ corresponds to the instant of DM scattering and phonon creation.}
\label{fig:J_geom}
\end{figure}

Next, now that the spatial dimension can be directly included in our simulations, we study the effect of the location where the DM generates an optical phonon in the target, $r_0$ (or, equivalently, the distance from the collector, $d = L - r_0$). In Figure~\ref{fig:J_geom}, where we plot the power as a function of time right after the target/collector interface, we found that the onset of the signal trivially depends on the distance that the phonons must travel before they reach the interface, as can be clearly seen in the inset. On the other hand, these differences tend to blur out at long simulation times, when phonons are transmitted after multiple reflections within the target (i.e., between the target/collector interface and the target backsurface) and then small differences in the initial conditions are less important.
Nevertheless, the difference tends to decrease at long simulation times, thicker devices result in larger signal (e.g.\ the power in the case of $L=1000$~nm is always larger than the one for $L=10$ or 100~nm). This last can be seen as the direct consequence of the dominance of boundary scattering in thinner devices,  which as harmonic, do not allow for decaying to transmissible states, thus reducing the overall collected signal.% (see Figure~\ref{fig:J_geom}).

We recall that, due to the slow response time of TES, this is the important region for our analysis. Notice that in these calculations we assumed Al to be oriented along the $\langle 111 \rangle$ axis (see also the sketch in Figure~\ref{fig:mc_sketch}); this matters when it comes to calculate the TBR. To make sure that no dramatic changes arise for different interfaces, we have repeated the calculations considering the $\langle 110 \rangle$ crystallographic orientation for the Al collector, finding that changes are negligible (see Supplementary Figure 1).

% Next, we studied the effect of the target volume. The same initial
%phonon state were time propagated in samples of 50~cm$^3$, 9~nm$^3$, 
%and 6~nm$^3$. A phonon of a given energy represents a much larger
%perturbation in a smaller volume and one can in principle expect
%the resulting dynamics to be altered. Contrary to this, however, 
%we do not detect any significant difference, as illustrated in 
%Figure~\ref{fig:volume}. In particular, all the conclusions drawn
%above about the sensitivity to energy and momentum remain valid.

Finally, we present what is perhaps the most striking result of
our calculations. The common wisdom is that the mass disorder introduced by a given isotope distribution is a source of scattering and should result in lower heat fluxes and thus power. Indeed, one of the reasons for selecting \al2o3\ is its small isotope dispersion (Al has only one stable isotope, while 99.8\% O atoms are $^{16}$O) and a valid question is if one should not rather pick a naturally isotopically-pure semiconductor, e.g.\ AlP, as a target material or think about isotope enrichment. For instance, Harrelson and coworkers~\cite{HarrelsonArXiv21} found that isotopic scattering in GaAs
and Si-based targets is the dominant scattering process, with associated lifetimes that are significantly smaller than anharmonic ones. They concluded that this effect is mitigated only with isotopic enrichment greater than 99.9\%.

\begin{figure}[t]
\includegraphics[width=1.0\linewidth]{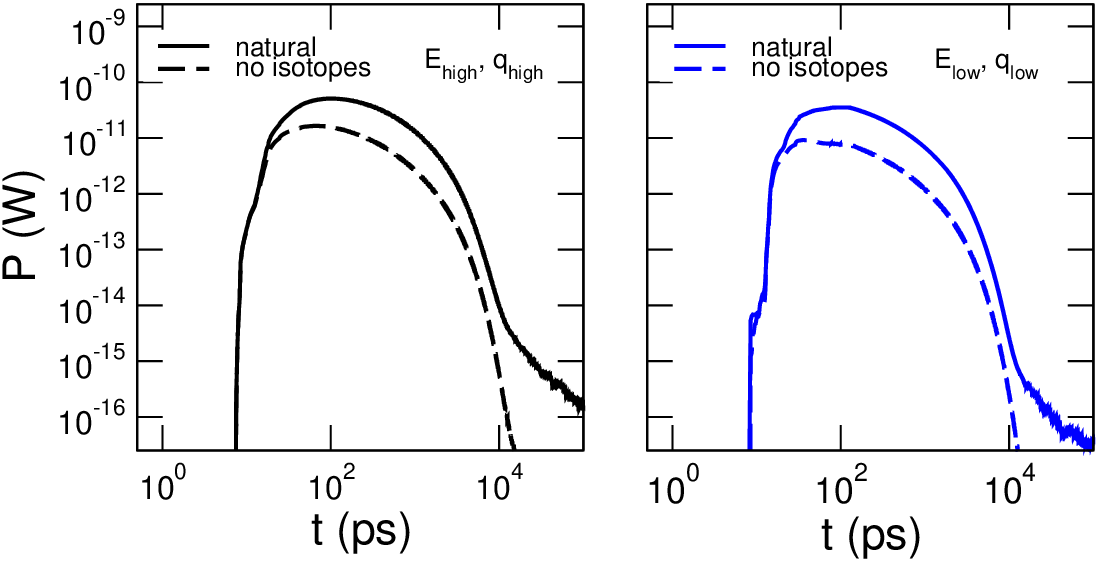}
       \caption{MC-computed power at the collector side (Al) of the interface as function of time, for the geometry shown in Fig.~\ref{fig:mc_sketch}, with (solid) and without (dashed) isotopic scattering for a high-energy-momentum (left) and a low-energy-momentum (right) initial mode. Those initial states have a quasi-momentum parallel to the \al2o3-$\langle0001\rangle$ direction. The initial phonon is in the same relative position ($r_0/L$) and $t=0$ corresponds to the instant of DM scattering and phonon creation.} 
\label{fig:J_iso}
\end{figure}

At variance with these considerations, however,
% we found a more complex scenario, where the effect of isotopes on the heat flux depend on the energy of the excited phonon. In the case of low energy phonons, the presencie of isotopes results in a lower signal, as expected. Conversely, for high energy phonons
we found that the main role of isotopes is providing, via scattering,  transport channels of higher velocities. This is shown in Figure~\ref{fig:J_iso}, where we compare the time evolution of the heat power in natural and isotopically purified \al2o3 (i.e.\ all O atoms are $^{16}$O). As it can be seen, the behavior is qualitatively similar, regardless of the values of the energy and momentum of the initially excited phonon: switching off isotope scattering {\it decreases}, rather than increases, the heat flux. A reliable estimate of this decrease at large values of $t$ is difficult to make, because the signal of the isotopically-pure case is so low that it is comparable to MC statistical noise. However, we conservatively estimate such a reduction to be of around two orders of magnitude for $t \approx 100$~ns. It is difficult to generalize this result to other materials, as it ultimately depends on the isotope population, isotope scattering rates and phonon velocity distribution. Nevertheless, at least in the case of \al2o3, rather than a hurdle, isotopes can even become an ally when it comes to maximizing the signal that reaches the superconducting film.
% \textcolor{cyan}{LF: THIS IS A VERY IMPORTANT RESULT WHICH WE DESCRIBE VERY SHORTLY. EVEN IF IT IS NOT CONVENIENT TO CHANGE FIG.10 TO LOG-LIN SCALE: CAN WE PROVIDE (WRITE) A NUMBER FOR THE FACTOR OF CHANGE OF THE HEAT FLOW DUE TO ISOTOPE SCATTERING AT 100ns, FOR INSTANCE? TO GIVE AN IDEA OF THE SIGNAL INCREASE DUE TO ISOTOPE EFFECTS. IN THE GRAPH THE ISOTOPE-FREE SIGNAL IS MISSED AT LONG TIMES.}

\smallskip
\noindent
{\bf Daily modulation and phonon dynamics.}~As Earth rotates the alignment of the detector with the dark matter galactic wind changes; so that phonons with different momenta---i.e.\ aligned to the dark matter wind orientation---are preferentially generated during different times of the day. Moreover, as the DM interaction is bound to the crystalline symmetry through the collective excitations in the solid (phonons), two nonequivalent crystalline directions will interact differently with the DM. Consequently, for anisotropic crystals, a daily modulation of the signal is expected, due to the different interaction strengths for the wind-oriented initial states. This particular feature has been proposed as a way to discern a DM generated signal from other sources and/or noise; nevertheless the effect and possible masking by downconversion processes of such an interesting feature for detection has never been discussed.

In Supplementary Figures~2-3 we present the signal for initial states with momentum parallel and perpendicular to the $\langle0001\rangle$ direction, which correspond respectively to the ordinary and extraordinary DM interaction axis of \al2o3~\cite{Knapen:2021bwg}.
It is clear from those, that phonon downconversion together with interface scattering yield similar fluxes for both---that is perpendicular and parallel---initial states. Therefore, any initial imbalance in the phonon generation rate would directly translate to the collected flux. In other words, the daily modulation in the Al\textsubscript{2}O\textsubscript{3}-$\langle111\rangle$ Al scheme is not masked either by phonon dynamics within the target or by the target-collector interface scattering.

\begin{figure}[t]
\includegraphics[width=1.0\linewidth]{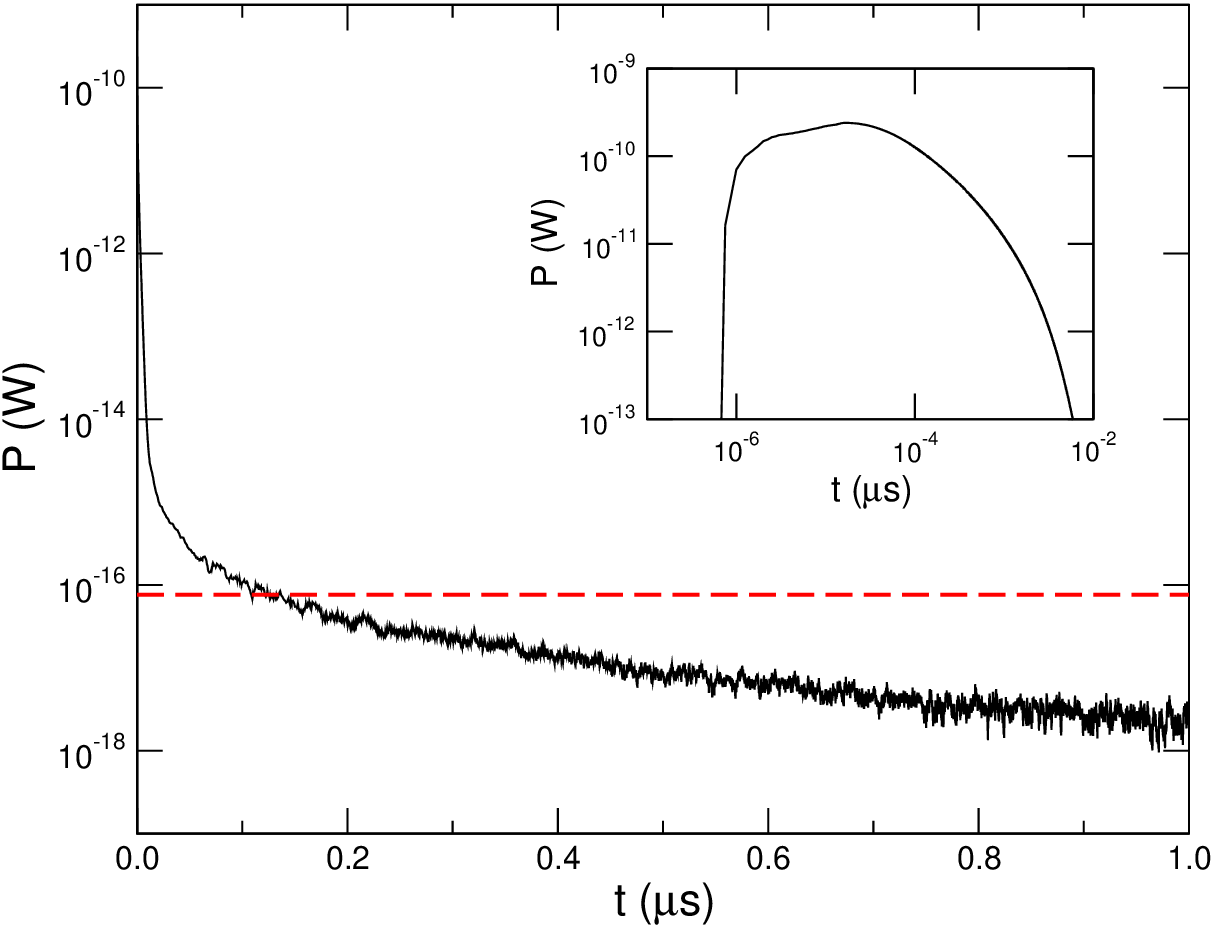}
       \caption{MC-computed power at the collector side (Al) of the interface as function of time. The red slashed line indicates the detectability limit calculated from the noise equivalent power (NEP) and bandwidth of a state-of-the-art TES device~\cite{FinkAIPAdv20, FinkAPL21}.
       Inset: log-log plot highlighting the peak of the power at very short values of $t$.} 
\label{fig:power}
\end{figure}

\smallskip
\noindent
{\bf Feasibility of phonon-based detection of sub-MeV DM.}~Up to this point we have studied the influence of various parameters --the phonon energy, the phonon momentum, the presence of isotopes, where the athermal phonon is generated within the target-- on the signal that reaches the collector and we have obtained a set of design rules for an efficient detector. Now we assess the overall feasibility of the detection scheme considered. In other words, can a carefully optimized detector provide a sufficiently intense signal that a TES can detect?

To this end, we have extended one of our Monte Carlo simulations up to 1~$\mu$s, focusing on the Al\textsubscript{3}O\textsubscript{2}$\langle0001\rangle$/Al$\langle111\rangle$ case with $L=\SI{10}{\nano\meter}$, $E=\SI{0.06}{\electronvolt}$, $\lVert\mathbf{q}\rVert=\SI{1}{eV}$, and $d=\SI{2.5}{\nano\meter}$. Then we plotted the power as a function of time and compared to the sensitivity  of state-of-the-art TESs (we have assumed $\mathrm{NEP} = 1.5\times 10^{-18}\,\mathrm{W}/\sqrt{\mathrm{Hz}}$ and $\mathrm{bandwidth} = 2.6 \,\mathrm{kHz}$)~\cite{FinkAIPAdv20,FinkAPL21}. The results are shown in Figure~\ref{fig:power}. As mentioned above, TES have rather slow response times, $t_r \gtrsim \mu\mathrm{s}$, and this plot highlights how critical this parameter is. The signal would be in the limit of detectability for state-of-the-art single TES devices ~\cite{FinkAIPAdv20, Khosropanah10}, indicating the need for further improvement of both time response and NEP of TES. The detection scheme, though, seems feasible either with the progress of present sensors or with new designs~\cite{PaolucciJAP20}. 

%BJK: The sensitivity estimates here are a little more complicated: https://arxiv.org/abs/2301.07617
A related question is whether sufficient target volume can be achieved to be sensitive to well-motivated models of Dark Matter. Many studies have provided a simple estimate of the sensitivity by calculating the exposure which is required in order to produce $\mathcal{O}(1)$ DM-induced phonons~\cite{KnapenPLB18,GriffinPRD18,Campbell-Deem:2019hdx,GriffinPRD20,Griffin:2020lgd,Kahn:2020fef,Trickle:2020oki,Coskuner:2021qxo,Campbell-Deem:2022fqm}. Calculating the DM-phonon scattering rate using \shell{DarkELF}~\cite{Knapen:2021bwg}, we find that currently unconstrained parameter space can be explore with an exposure $\sim 10^{-3}\,\text{g-days}$, while models in which DM is produced with the correct relic abundance via freeze-in would require an exposure of $\sim 1\,\text{g-days}$~\cite{GriffinPRD18}. Considering a few days of exposure and a surface area of $\sim 100\,\mathrm{cm}^2$ for the target crystal, this would require target thicknesses of $\mathcal{O}(10\,\mathrm{nm} - 10\,\mu\mathrm{m})$.
Of course, a detailed calculation is required, including the phonon energy losses described in this work, but this estimate demonstrates that wafers of realistic thicknesses can be a feasible target for probing unexplored DM parameter space.

Finally, we find it important to highlight that our results should be understood as an upper limit, as a real device will contain further sources of signal diminution. These include sample defects or the partial coverage of the target by the collector that has not been considered in our simulations, not to forget about losses related to other processes within the collector that are not related with phonon dynamics. Also, our simulation scheme contains several approximations to reduce the computational burden (e.g. use of bulk properties, elastic scattering at the interface, etc.) that can further affect/diminish the signal (see Appendix~\ref{app:limits} for a comprehensive list and discussion of the approximations and limitations of the applied methodology).

\section{Conclusions}
\label{sec:conc}

We have presented a detailed scrutiny of the phonon dynamics resulting from an athermal optical phonon excitation, followed by downconversion and propagation within a semiconductor target, until reaching the interface of a phonon collector. We consider these processes in the context of the detection of light DM, though they are general to a wider class of instruments based on cryogenic thermal detectors. Our formalism is entirely based on parameter-free, {\it ab initio} electronic structure calculations. We focused on an Al$_2$O$_3$/Al target/collector, two material systems that offer several advantages for the detection of light DM, such as a phonon spectrum of the target in the right energy range, an anisotropic response suited to reveal daily modulation and a high-quality interface between the semiconductor and the superconductor. 

A preliminary methodological survey revealed that any RTA-based approach is bound to yield predictions of limited accuracy, largely underestimating the heat flux that reaches the collector and then artificially dictating experimental conditions that are more difficult to meet. Conversely, linearization of the PBTE works well. Based on these conclusions, we developed a 3D beyond-RTA phonon Monte Carlo that allowed us to introduce the spatial dimension of the device and address questions about its size or the effect of where in the target the DM scattering takes place (i.e., the distance from the collector). We showed that the wavevector of the optical phonon initially excited by the DM particle plays a negligible role, while a higher energy phonon results in a (slightly) larger heat flux reaching the collector. Isotopes can, perhaps counterintuitively, favor the detection and result in a larger heat flux by providing transport channels of higher velocities. Analyzing these effects in materials with a larger dispersion in the isotope population (e.g., SiC) deserves a separate study to see if a more conventional behavior is recovered (i.e., isotopes decrease the heat flux) or if the behavior reported here is amplified. The mismatch in the phonon spectra of Al$_2$O$_3$ and Al is the source of a considerable thermal boundary resistance, which nonetheless does not limit significantly the efficiency of the detector: phonons undergo multiple reflections between the target/collector interface and the target backsurface, until they {\it fit} the narrower energy window of the collector. For this reason, too thick targets should be avoided (the multiple reflections would imply too large distances travelled, with the consequent losses), while the initial position of the phonon excitation does not seem to play a major role.

Altogether, our results suggest that, though challenging---particularly in relation to the sensitivity of the TES with respect to the collected power---the direct detection of light DM via athermal phonon generation using a TES-based detector appears to be feasible, or at least, the phonon downconversion followed by quasi-ballistic phonon propagation does not appear to be a major bottleneck in terms of reducing the signal.
We also stress that most of these conclusions are general and can be extended to any phonon-based cryogenic detection scheme, regardless of the source (DM in our case) of the initial phonon excitation. On the downside, many of the results presented here are expected to be material-dependent and thus the quantitative importance of the different factors should be carefully evaluated case by case.

\section*{ACKNOWLEDGMENTS}
We acknowledge financial support from AEI under grant PID2020-119777GB-I00, and the Severo Ochoa Centres of
Excellence Program under grant CEX2019-000917-S, by the Generalitat
de Catalunya under grant no. and 2017 SGR 1506, and by the Spanish NRRP 
through the CSIC PTI QTEP+. BJK acknowledges funding from the Ram\'on y Cajal Grant RYC2021-034757-I, financed by MCIN/AEI/10.13039/501100011033 and by the European Union ``NextGenerationEU"/PRTR. 
Calculations were performed at the
Centro de Supercomputaci\'on de Galicia (CESGA) within action
FI-2022-3-0031,``Phonon-based detection of ultralight dark matter'', of the Red Espa\~nola de Supercomputaci\'on (RES).
The authors also thank the `Dark Collaboration at IFCA' working group for useful discussions.

\section*{APPENDIX: METHODS EXTENDED}
\label{sec:Methods}
\subsection{Solving the Swartz's model}

The key ingredients of Swartz's model (Eq.~\ref{eq:GrossFlux}) are the time evolution of the phonon deviational population [$n^d(t)$], and the transmission probability ($\alpha$)
The former, under the assumption of homogeneity---as it is the case of Swartz's model---, can be obtained by solely time integrating the collision operator:
\begin{widetext}
\begin{multline}
	\label{eq:fullscattering}
	\left.\frac{\partial n_i}{\partial t} \right|_\mathrm{collision} =  -\sum_{jk}P^{\mathrm{3ph}}_{i+j\rightarrow k}[ n_i n_j (n_k+1) - 
	(n_i+1)(n_j+1)n_k] - \frac{1}{2}\sum_{jk}P^{\mathrm{3ph}}_{i\rightarrow j+k}[ n_i (n_j+1) (n_k+1) - (n_i+1)n_j n_k] - \\ \sum_{j}P^{\mathrm{iso}}_{i\rightarrow j}[ n_i (n_j+1) - (n_i+1)n_j] = 
	\sum_{jk}P^{\mathrm{3ph}}_{i+j\rightarrow k}[  (n^0_k-n^0_j) n^d_i + (n^0_k-n^0_i) n^d_j + (n^0_i+n^0_j+1) n^d_k   - \\ n_i^d n_j^d + n_i^d n_k^d + n_j^d n_k^d -n_i^0 n_j^0  + n^0_k (n_i^0 + n_j^0 + 1)] 
	+\frac{1}{2}\sum_{jk}P^{\mathrm{3ph}}_{i\rightarrow j+k}[ -(n^0_j+n_k^0+1)n^d_i + (n_k^0-n_j^0)n^d_j + (n_j^0-n^0_i)n^d_k - \\
	 n^d_i n^d_j - n_i^d n_k^d + n_j^d n_k^d -n^0_i (n^0_j+n^0_k+1) + n^0_j n^0_k] 
	+ \sum_{j}P^{\mathrm{iso}}_{i\rightarrow j}[
	-n_i^d + n_j^d  -n^0_i + n^0_j] ,
\end{multline}
\end{widetext}
and/or its variations, i.e. the linearized scattering operator---that is obtained by neglecting the order-2 deviational terms---and the pseudo-RTA (see Eq.~\ref{eq:pseudoRTA}). We note that Eq.~\ref{eq:fullscattering} only includes isotopic and three-phonon scattering process.

Generally, this equation contains two or more time scales of a different order, i.e. some phonon modes are evolving much faster than others, a characteristic typical of a stiff system. So that, the optimal methodology to integrate Eq.~\ref{eq:fullscattering} would be to make use of an implicit method (e.g. Rosenbrock methods or implicit Runge-Kutta methods); nonetheless, such methods require the computation of the Jacobian. However, the computation of such a matrix is computationally prohibitive in our case; for reference, in cubic GaAs, taking a $24\times24\times24$ $\mathbf{q}$-mesh would mean a Jacobian of $\sim$55 GB. Consequently, one is constricted to make use of explicit methods with a small enough timestep ($\Delta t$)---either arbitrarily selected by the user or chosen based on the RTA lifetimes, namely $\Delta t = \min\{0.25\mathrm{E}(\tau_i|\tau_i \neq 0),1\}$\SI{}{\pico\second}---to integrate $n_i(t)$.  

Amid all the available explicit methods, the predictor-corrector methods are especially suited for problems in which the computation of the derivative is computationally expensive, as it is the case of the full collision operator. Therefore, we have selected a two-step predictor-corrector method with a constant timestep as implemented in boost::odeint~\cite{boostodeint} library, to integrate Eq.~\ref{eq:fullscattering} and its variations, using an explicit $4^{\mathrm{th}}$ order Runge-Kutta to obtain the initial states required for the selected scheme. We note, that for the pseudo-RTA case it is much more efficient to directly solve the Eq.~\ref{eq:collisionAnd}; so that the deviational phonon population evolves accordingly to 
\begin{equation}
	n^d(t+\Delta t) = \exp(A \Delta t) n^d(t) \approx (I + A \Delta t) n^d(t),
\end{equation}
where we have expanded the exponential matrix [$\exp(A \Delta t)$] using a Taylor series up to the 1\textsuperscript{st} order. 

In the building of this collision operator it is essential for it to respect crystal symmetries and microscopic reversibility~\cite{LandonJAP14}; consequently we have used the symmetric adaptive smearing scheme of Ref.~\cite{Raya-MorenoCPC22} to compute the three-phonon processes.

Besides symmetry, it must also be noted that the linearized and the full scattering operator must in principle conserve the energy (note again here that RTA, and pseudo-RTA are energy destructive by construction); however, this condition is violated due to the broadening scheme, in which delta conservation energies are approximated by Gaussians. Therefore, we add a correction ($\beta_i$) extracted from a Lagrange-multiplier approach, similarly to the one developed in Ref.~\cite{LandonJAP14} to our derivative computation each step to enforce energy conservation. So that, given a violation of energy, $\Delta E = \sum_i \hbar \omega_i\frac{\partial n^d_i}{\partial t}$, imposes a constraint for the applied correction $\left(	\sum_i \beta_i = -\Delta E \right)$.  Moreover, in order to reduce possible side effects of such a correction, $\beta_i$ must remain as small as possible. This can be achieved through a Lagrange multiplier's through the optimization of the following constrained objective function:
\begin{equation}
F = \sum_i \beta_i^2 + \lambda\left(\sum_i \beta_i  + \Delta E\right),
\end{equation}
and thus,
\begin{gather}
	\frac{\partial F}{\partial \beta_j} = 2\beta_j + \lambda , 
\end{gather}
which is minimized by $\beta_j = -\frac{\lambda}{2}$. Consequently, combining this last with the constraint, one arrives to: 
\begin{equation}
	 \beta_i = -\frac{\sum_j \hbar \omega_j\frac{\partial n^d_j}{\partial t}}{N_{\mathrm{modes}}}
\end{equation}
where ${N_{\mathrm{modes}}}$ is the number of modes.

Finally, as for the latter ingredient, the transmission coefficient, we compute it using as previously mentioned the DMM; nevertheless we also allow for a uniform setting of an equal transmission coefficient for all the modes. This last is a very simplistic model that works for pairs with similar maximum frequencies like GaAs-Si, but yields unreasonable results for most cases, like the present case, as it would allow transmission from high-energy modes of \al2o3 that do not exist in Al.

\subsection{Practical notes on MC implementation: the propagator}

The basics of the MC solver can be found on Ref.~\cite{Raya-MorenoCPC22}---\texttt{BTE-Barna}, which this code is derived from---, and more theoreticall background can be found in Refs.~\cite{LandonJAP14} and \cite{LandonPHDTHESIS14}. Here we only highlight the main difference with respect to \texttt{BTE-Barna}, that is the computation of the propagator, namely $\exp(B\delta t)$, required for the scattering algorithm~\cite{LandonPHDTHESIS14}. In contrast with previous works~\cite{Raya-MorenoCPC22,LandonJAP14,LandonPHDTHESIS14}, we are dealing with 3D materials instead of 2D ones, rocketing the memory and computational resources required for the computation; therefore, it becomes desirable to keep up not only the sparsity of the transition matrix but in the propagator itself. 

Consequently, we introduce the energy conservation correction only on non-zero entries, so that the transition matrix sparsity is conserved, but we also make use of the first-order Taylor expansion to the exponential---i.e. $\exp(B\delta t) \approx I + B\delta t$ to translate this sparsity to the propagator itself. Nonetheless, this last approximation can seem crude, it was proven by Landon that for small time steps---namely like those used to solve the Swartz's model---is virtually equivalent to more refined methods~\cite{LandonPHDTHESIS14}.

\subsection{\label{app:limits}Practical notes on MC implementation: approximations and limitations}

Although our approach is in principle the most exact one among those used up to now, our simulations have still several limits and approximations that should be kept on mind, most of them coming from a design point of view. Here we highlight some of these limitations.
\begin{itemize}
    \item \textbf{Boundary scattering:} The implementation is like the one of the RTA MC solvers included in \texttt{BTE-Barna}, i.e. perfectly elastic and diffusive.
    \item \textbf{Spatial resolution on in-plane direction:} The in-plane spatial discretitation is disregarded, and the system is assumed to be infinite in those directions. This last can be only considered if the transport direction (out-of-plane) is much smaller than the others.
    \item \textbf{Phonons in the collector:} we are assuming that phonons do not reenter from the collector to the target. This last is partially justified by them being absorbed in breaking the Cooper pairs in the collector. 
    \item \textbf{Phonon properties in the target/collector:} we are assuming bulk phonon properties, both harmonic and anharmonic in both target and collector. 
\end{itemize}

% \bibliography{complete}

\begin{thebibliography}{79}%
\makeatletter
\providecommand \@ifxundefined [1]{%
 \@ifx{#1\undefined}
}%
\providecommand \@ifnum [1]{%
 \ifnum #1\expandafter \@firstoftwo
 \else \expandafter \@secondoftwo
 \fi
}%
\providecommand \@ifx [1]{%
 \ifx #1\expandafter \@firstoftwo
 \else \expandafter \@secondoftwo
 \fi
}%
\providecommand \natexlab [1]{#1}%
\providecommand \enquote  [1]{``#1''}%
\providecommand \bibnamefont  [1]{#1}%
\providecommand \bibfnamefont [1]{#1}%
\providecommand \citenamefont [1]{#1}%
\providecommand \href@noop [0]{\@secondoftwo}%
\providecommand \href [0]{\begingroup \@sanitize@url \@href}%
\providecommand \@href[1]{\@@startlink{#1}\@@href}%
\providecommand \@@href[1]{\endgroup#1\@@endlink}%
\providecommand \@sanitize@url [0]{\catcode `\\12\catcode `\$12\catcode
  `\&12\catcode `\#12\catcode `\^12\catcode `\_12\catcode `\%12\relax}%
\providecommand \@@startlink[1]{}%
\providecommand \@@endlink[0]{}%
\providecommand \url  [0]{\begingroup\@sanitize@url \@url }%
\providecommand \@url [1]{\endgroup\@href {#1}{\urlprefix }}%
\providecommand \urlprefix  [0]{URL }%
\providecommand \Eprint [0]{\href }%
\providecommand \doibase [0]{https://doi.org/}%
\providecommand \selectlanguage [0]{\@gobble}%
\providecommand \bibinfo  [0]{\@secondoftwo}%
\providecommand \bibfield  [0]{\@secondoftwo}%
\providecommand \translation [1]{[#1]}%
\providecommand \BibitemOpen [0]{}%
\providecommand \bibitemStop [0]{}%
\providecommand \bibitemNoStop [0]{.\EOS\space}%
\providecommand \EOS [0]{\spacefactor3000\relax}%
\providecommand \BibitemShut  [1]{\csname bibitem#1\endcsname}%
\let\auto@bib@innerbib\@empty
%</preamble>
\bibitem [{\citenamefont {Bertone}\ \emph {et~al.}(2005)\citenamefont
  {Bertone}, \citenamefont {Hooper},\ and\ \citenamefont
  {Silk}}]{Bertone:2004pz}%
  \BibitemOpen
  \bibfield  {author} {\bibinfo {author} {\bibfnamefont {G.}~\bibnamefont
  {Bertone}}, \bibinfo {author} {\bibfnamefont {D.}~\bibnamefont {Hooper}},\
  and\ \bibinfo {author} {\bibfnamefont {J.}~\bibnamefont {Silk}},\ }\href
  {https://doi.org/10.1016/j.physrep.2004.08.031} {\bibfield  {journal}
  {\bibinfo  {journal} {Phys. Rept.}\ }\textbf {\bibinfo {volume} {405}},\
  \bibinfo {pages} {279} (\bibinfo {year} {2005})},\ \Eprint
  {https://arxiv.org/abs/hep-ph/0404175} {arXiv:hep-ph/0404175} \BibitemShut
  {NoStop}%
\bibitem [{\citenamefont {Gaskins}(2016)}]{Gaskins:2016cha}%
  \BibitemOpen
  \bibfield  {author} {\bibinfo {author} {\bibfnamefont {J.~M.}\ \bibnamefont
  {Gaskins}},\ }\href {https://doi.org/10.1080/00107514.2016.1175160}
  {\bibfield  {journal} {\bibinfo  {journal} {Contemp. Phys.}\ }\textbf
  {\bibinfo {volume} {57}},\ \bibinfo {pages} {496} (\bibinfo {year} {2016})},\
  \Eprint {https://arxiv.org/abs/1604.00014} {arXiv:1604.00014 [astro-ph.HE]}
  \BibitemShut {NoStop}%
\bibitem [{\citenamefont {Boveia}\ and\ \citenamefont
  {Doglioni}(2018)}]{Boveia:2018yeb}%
  \BibitemOpen
  \bibfield  {author} {\bibinfo {author} {\bibfnamefont {A.}~\bibnamefont
  {Boveia}}\ and\ \bibinfo {author} {\bibfnamefont {C.}~\bibnamefont
  {Doglioni}},\ }\href {https://doi.org/10.1146/annurev-nucl-101917-021008}
  {\bibfield  {journal} {\bibinfo  {journal} {Ann. Rev. Nucl. Part. Sci.}\
  }\textbf {\bibinfo {volume} {68}},\ \bibinfo {pages} {429} (\bibinfo {year}
  {2018})},\ \Eprint {https://arxiv.org/abs/1810.12238} {arXiv:1810.12238
  [hep-ex]} \BibitemShut {NoStop}%
\bibitem [{\citenamefont {Billard}\ \emph {et~al.}(2022)\citenamefont {Billard}
  \emph {et~al.}}]{Billard:2021uyg}%
  \BibitemOpen
  \bibfield  {author} {\bibinfo {author} {\bibfnamefont {J.}~\bibnamefont
  {Billard}} \emph {et~al.},\ }\href {https://doi.org/10.1088/1361-6633/ac5754}
  {\bibfield  {journal} {\bibinfo  {journal} {Rept. Prog. Phys.}\ }\textbf
  {\bibinfo {volume} {85}},\ \bibinfo {pages} {056201} (\bibinfo {year}
  {2022})},\ \Eprint {https://arxiv.org/abs/2104.07634} {arXiv:2104.07634
  [hep-ex]} \BibitemShut {NoStop}%
\bibitem [{\citenamefont {Bertone}\ and\ \citenamefont
  {Tait}(2018)}]{Bertone:2018krk}%
  \BibitemOpen
  \bibfield  {author} {\bibinfo {author} {\bibfnamefont {G.}~\bibnamefont
  {Bertone}}\ and\ \bibinfo {author} {\bibfnamefont {T.}~\bibnamefont {Tait},
  \bibfnamefont {M.~P.}},\ }\href {https://doi.org/10.1038/s41586-018-0542-z}
  {\bibfield  {journal} {\bibinfo  {journal} {Nature}\ }\textbf {\bibinfo
  {volume} {562}},\ \bibinfo {pages} {51} (\bibinfo {year} {2018})},\ \Eprint
  {https://arxiv.org/abs/1810.01668} {arXiv:1810.01668 [astro-ph.CO]}
  \BibitemShut {NoStop}%
\bibitem [{\citenamefont {Essig}\ \emph {et~al.}(2012)\citenamefont {Essig},
  \citenamefont {Mardon},\ and\ \citenamefont {Volansky}}]{Essig:2011nj}%
  \BibitemOpen
  \bibfield  {author} {\bibinfo {author} {\bibfnamefont {R.}~\bibnamefont
  {Essig}}, \bibinfo {author} {\bibfnamefont {J.}~\bibnamefont {Mardon}},\ and\
  \bibinfo {author} {\bibfnamefont {T.}~\bibnamefont {Volansky}},\ }\href
  {https://doi.org/10.1103/PhysRevD.85.076007} {\bibfield  {journal} {\bibinfo
  {journal} {Phys. Rev. D}\ }\textbf {\bibinfo {volume} {85}},\ \bibinfo
  {pages} {076007} (\bibinfo {year} {2012})},\ \Eprint
  {https://arxiv.org/abs/1108.5383} {arXiv:1108.5383 [hep-ph]} \BibitemShut
  {NoStop}%
\bibitem [{\citenamefont {Green}(2017)}]{Green:2017odb}%
  \BibitemOpen
  \bibfield  {author} {\bibinfo {author} {\bibfnamefont {A.~M.}\ \bibnamefont
  {Green}},\ }\href {https://doi.org/10.1088/1361-6471/aa7819} {\bibfield
  {journal} {\bibinfo  {journal} {J. Phys. G}\ }\textbf {\bibinfo {volume}
  {44}},\ \bibinfo {pages} {084001} (\bibinfo {year} {2017})},\ \Eprint
  {https://arxiv.org/abs/1703.10102} {arXiv:1703.10102 [astro-ph.CO]}
  \BibitemShut {NoStop}%
\bibitem [{\citenamefont {Knapen}\ \emph {et~al.}(2018)\citenamefont {Knapen},
  \citenamefont {Lin}, \citenamefont {Pyle},\ and\ \citenamefont
  {Zurek}}]{KnapenPLB18}%
  \BibitemOpen
  \bibfield  {author} {\bibinfo {author} {\bibfnamefont {S.}~\bibnamefont
  {Knapen}}, \bibinfo {author} {\bibfnamefont {T.}~\bibnamefont {Lin}},
  \bibinfo {author} {\bibfnamefont {M.}~\bibnamefont {Pyle}},\ and\ \bibinfo
  {author} {\bibfnamefont {K.~M.}\ \bibnamefont {Zurek}},\ }\href
  {https://doi.org/10.1016/j.physletb.2018.08.064} {\bibfield  {journal}
  {\bibinfo  {journal} {Phys. Lett. B}\ }\textbf {\bibinfo {volume} {785}},\
  \bibinfo {pages} {386} (\bibinfo {year} {2018})},\ \Eprint
  {https://arxiv.org/abs/1712.06598} {arXiv:1712.06598 [hep-ph]} \BibitemShut
  {NoStop}%
\bibitem [{\citenamefont {Chigusa}\ \emph {et~al.}(2020)\citenamefont
  {Chigusa}, \citenamefont {Moroi},\ and\ \citenamefont
  {Nakayama}}]{ChigusaPRD20}%
  \BibitemOpen
  \bibfield  {author} {\bibinfo {author} {\bibfnamefont {S.}~\bibnamefont
  {Chigusa}}, \bibinfo {author} {\bibfnamefont {T.}~\bibnamefont {Moroi}},\
  and\ \bibinfo {author} {\bibfnamefont {K.}~\bibnamefont {Nakayama}},\
  }\href@noop {} {\bibfield  {journal} {\bibinfo  {journal} {Phys. Rev. D}\
  }\textbf {\bibinfo {volume} {101}},\ \bibinfo {pages} {096013} (\bibinfo
  {year} {2020})}\BibitemShut {NoStop}%
\bibitem [{\citenamefont {Mitridate}\ \emph {et~al.}(2020)\citenamefont
  {Mitridate}, \citenamefont {Trickle}, \citenamefont {Zhang},\ and\
  \citenamefont {Zurek}}]{MitridatePRD20}%
  \BibitemOpen
  \bibfield  {author} {\bibinfo {author} {\bibfnamefont {A.}~\bibnamefont
  {Mitridate}}, \bibinfo {author} {\bibfnamefont {T.}~\bibnamefont {Trickle}},
  \bibinfo {author} {\bibfnamefont {Z.}~\bibnamefont {Zhang}},\ and\ \bibinfo
  {author} {\bibfnamefont {K.~M.}\ \bibnamefont {Zurek}},\ }\href
  {https://doi.org/10.1103/PhysRevD.102.095005} {\bibfield  {journal} {\bibinfo
   {journal} {Phys. Rev. D}\ }\textbf {\bibinfo {volume} {102}},\ \bibinfo
  {pages} {095005} (\bibinfo {year} {2020})},\ \Eprint
  {https://arxiv.org/abs/2005.10256} {arXiv:2005.10256 [hep-ph]} \BibitemShut
  {NoStop}%
\bibitem [{\citenamefont {Hochberg}\ \emph {et~al.}(2017)\citenamefont
  {Hochberg}, \citenamefont {Lin},\ and\ \citenamefont
  {Zurek}}]{HochbergPRD17}%
  \BibitemOpen
  \bibfield  {author} {\bibinfo {author} {\bibfnamefont {Y.}~\bibnamefont
  {Hochberg}}, \bibinfo {author} {\bibfnamefont {T.}~\bibnamefont {Lin}},\ and\
  \bibinfo {author} {\bibfnamefont {K.~M.}\ \bibnamefont {Zurek}},\ }\href
  {https://doi.org/10.1103/PhysRevD.95.023013} {\bibfield  {journal} {\bibinfo
  {journal} {Phys. Rev. D}\ }\textbf {\bibinfo {volume} {95}},\ \bibinfo
  {pages} {023013} (\bibinfo {year} {2017})},\ \Eprint
  {https://arxiv.org/abs/1608.01994} {arXiv:1608.01994 [hep-ph]} \BibitemShut
  {NoStop}%
\bibitem [{\citenamefont {Bloch}\ \emph {et~al.}(2017)\citenamefont {Bloch},
  \citenamefont {Essig}, \citenamefont {Tobioka}, \citenamefont {Volansky},\
  and\ \citenamefont {Yu}}]{BlochJHEP17}%
  \BibitemOpen
  \bibfield  {author} {\bibinfo {author} {\bibfnamefont {I.~M.}\ \bibnamefont
  {Bloch}}, \bibinfo {author} {\bibfnamefont {R.}~\bibnamefont {Essig}},
  \bibinfo {author} {\bibfnamefont {K.}~\bibnamefont {Tobioka}}, \bibinfo
  {author} {\bibfnamefont {T.}~\bibnamefont {Volansky}},\ and\ \bibinfo
  {author} {\bibfnamefont {T.-T.}\ \bibnamefont {Yu}},\ }\href
  {https://doi.org/10.1007/JHEP06(2017)087} {\bibfield  {journal} {\bibinfo
  {journal} {JHEP}\ }\textbf {\bibinfo {volume} {06}},\ \bibinfo {pages}
  {087}},\ \Eprint {https://arxiv.org/abs/1608.02123} {arXiv:1608.02123
  [hep-ph]} \BibitemShut {NoStop}%
\bibitem [{\citenamefont {Hochberg}\ \emph {et~al.}(2016)\citenamefont
  {Hochberg}, \citenamefont {Lin},\ and\ \citenamefont
  {Zurek}}]{Hochberg:2016ajh}%
  \BibitemOpen
  \bibfield  {author} {\bibinfo {author} {\bibfnamefont {Y.}~\bibnamefont
  {Hochberg}}, \bibinfo {author} {\bibfnamefont {T.}~\bibnamefont {Lin}},\ and\
  \bibinfo {author} {\bibfnamefont {K.~M.}\ \bibnamefont {Zurek}},\ }\href
  {https://doi.org/10.1103/PhysRevD.94.015019} {\bibfield  {journal} {\bibinfo
  {journal} {Phys. Rev. D}\ }\textbf {\bibinfo {volume} {94}},\ \bibinfo
  {pages} {015019} (\bibinfo {year} {2016})},\ \Eprint
  {https://arxiv.org/abs/1604.06800} {arXiv:1604.06800 [hep-ph]} \BibitemShut
  {NoStop}%
\bibitem [{\citenamefont {Hochberg}\ \emph {et~al.}(2019)\citenamefont
  {Hochberg}, \citenamefont {Charaev}, \citenamefont {Nam}, \citenamefont
  {Verma}, \citenamefont {Colangelo},\ and\ \citenamefont
  {Berggren}}]{Hochberg:2019cyy}%
  \BibitemOpen
  \bibfield  {author} {\bibinfo {author} {\bibfnamefont {Y.}~\bibnamefont
  {Hochberg}}, \bibinfo {author} {\bibfnamefont {I.}~\bibnamefont {Charaev}},
  \bibinfo {author} {\bibfnamefont {S.-W.}\ \bibnamefont {Nam}}, \bibinfo
  {author} {\bibfnamefont {V.}~\bibnamefont {Verma}}, \bibinfo {author}
  {\bibfnamefont {M.}~\bibnamefont {Colangelo}},\ and\ \bibinfo {author}
  {\bibfnamefont {K.~K.}\ \bibnamefont {Berggren}},\ }\href
  {https://doi.org/10.1103/PhysRevLett.123.151802} {\bibfield  {journal}
  {\bibinfo  {journal} {Phys. Rev. Lett.}\ }\textbf {\bibinfo {volume} {123}},\
  \bibinfo {pages} {151802} (\bibinfo {year} {2019})},\ \Eprint
  {https://arxiv.org/abs/1903.05101} {arXiv:1903.05101 [hep-ph]} \BibitemShut
  {NoStop}%
\bibitem [{\citenamefont {Griffin}\ \emph {et~al.}(2018)\citenamefont
  {Griffin}, \citenamefont {Knapen}, \citenamefont {Lin},\ and\ \citenamefont
  {Zurek}}]{GriffinPRD18}%
  \BibitemOpen
  \bibfield  {author} {\bibinfo {author} {\bibfnamefont {S.}~\bibnamefont
  {Griffin}}, \bibinfo {author} {\bibfnamefont {S.}~\bibnamefont {Knapen}},
  \bibinfo {author} {\bibfnamefont {T.}~\bibnamefont {Lin}},\ and\ \bibinfo
  {author} {\bibfnamefont {K.~M.}\ \bibnamefont {Zurek}},\ }\href
  {https://doi.org/10.1103/PhysRevD.98.115034} {\bibfield  {journal} {\bibinfo
  {journal} {Phys. Rev. D}\ }\textbf {\bibinfo {volume} {98}},\ \bibinfo
  {pages} {115034} (\bibinfo {year} {2018})},\ \Eprint
  {https://arxiv.org/abs/1807.10291} {arXiv:1807.10291 [hep-ph]} \BibitemShut
  {NoStop}%
\bibitem [{\citenamefont {Pirro}\ and\ \citenamefont
  {Mauskopf}(2017)}]{PirroARNPS17}%
  \BibitemOpen
  \bibfield  {author} {\bibinfo {author} {\bibfnamefont {S.}~\bibnamefont
  {Pirro}}\ and\ \bibinfo {author} {\bibfnamefont {P.}~\bibnamefont
  {Mauskopf}},\ }\href {https://doi.org/10.1146/annurev-nucl-101916-123130}
  {\bibfield  {journal} {\bibinfo  {journal} {Ann. Rev. Nucl. Part. Sci.}\
  }\textbf {\bibinfo {volume} {67}},\ \bibinfo {pages} {161} (\bibinfo {year}
  {2017})}\BibitemShut {NoStop}%
\bibitem [{\citenamefont {Irwin}\ and\ \citenamefont
  {Hilton}(2005)}]{IrwinTAP07}%
  \BibitemOpen
  \bibfield  {author} {\bibinfo {author} {\bibfnamefont {K.~D.}\ \bibnamefont
  {Irwin}}\ and\ \bibinfo {author} {\bibfnamefont {G.~C.}\ \bibnamefont
  {Hilton}},\ }in\ \href@noop {} {\emph {\bibinfo {booktitle} {Topics in Appl.
  Phys.}}},\ Vol.~\bibinfo {volume} {99},\ \bibinfo {editor} {edited by\
  \bibinfo {editor} {\bibfnamefont {C.}~\bibnamefont {Enss}}}\ (\bibinfo
  {publisher} {Springer Berlin, Heidelberg},\ \bibinfo {year}
  {2005})\BibitemShut {NoStop}%
\bibitem [{\citenamefont {Angloher}\ \emph {et~al.}(2005)\citenamefont
  {Angloher} \emph {et~al.}}]{AngloherAP05}%
  \BibitemOpen
  \bibfield  {author} {\bibinfo {author} {\bibfnamefont {G.}~\bibnamefont
  {Angloher}} \emph {et~al.},\ }\href
  {https://doi.org/10.1016/j.astropartphys.2005.01.006} {\bibfield  {journal}
  {\bibinfo  {journal} {Astropart. Phys.}\ }\textbf {\bibinfo {volume} {23}},\
  \bibinfo {pages} {325} (\bibinfo {year} {2005})},\ \Eprint
  {https://arxiv.org/abs/astro-ph/0408006} {arXiv:astro-ph/0408006}
  \BibitemShut {NoStop}%
\bibitem [{\citenamefont {Strauss}\ \emph {et~al.}(2017)\citenamefont {Strauss}
  \emph {et~al.}}]{StraussNIMA17}%
  \BibitemOpen
  \bibfield  {author} {\bibinfo {author} {\bibfnamefont {R.}~\bibnamefont
  {Strauss}} \emph {et~al.},\ }\href
  {https://doi.org/10.1016/j.nima.2016.06.060} {\bibfield  {journal} {\bibinfo
  {journal} {Nucl. Instrum. Meth. A}\ }\textbf {\bibinfo {volume} {845}},\
  \bibinfo {pages} {414} (\bibinfo {year} {2017})},\ \Eprint
  {https://arxiv.org/abs/1802.08639} {arXiv:1802.08639 [astro-ph.IM]}
  \BibitemShut {NoStop}%
\bibitem [{\citenamefont {Agnese}\ \emph {et~al.}(2014)\citenamefont {Agnese}
  \emph {et~al.}}]{AgnesePRL14}%
  \BibitemOpen
  \bibfield  {author} {\bibinfo {author} {\bibfnamefont {R.}~\bibnamefont
  {Agnese}} \emph {et~al.} (\bibinfo {collaboration} {SuperCDMS
  Collaboration}),\ }\href {https://doi.org/10.1103/PhysRevLett.112.041302}
  {\bibfield  {journal} {\bibinfo  {journal} {Phys. Rev. Lett.}\ }\textbf
  {\bibinfo {volume} {112}},\ \bibinfo {pages} {041302} (\bibinfo {year}
  {2014})},\ \Eprint {https://arxiv.org/abs/1309.3259} {arXiv:1309.3259
  [physics.ins-det]} \BibitemShut {NoStop}%
\bibitem [{\citenamefont {Agnese}\ \emph {et~al.}(2017)\citenamefont {Agnese}
  \emph {et~al.}}]{AgnesePRD17}%
  \BibitemOpen
  \bibfield  {author} {\bibinfo {author} {\bibfnamefont {R.}~\bibnamefont
  {Agnese}} \emph {et~al.} (\bibinfo {collaboration} {SuperCDMS
  Collaboration}),\ }\href {https://doi.org/10.1103/PhysRevD.95.082002}
  {\bibfield  {journal} {\bibinfo  {journal} {Phys. Rev. D}\ }\textbf {\bibinfo
  {volume} {95}},\ \bibinfo {pages} {082002} (\bibinfo {year} {2017})},\
  \Eprint {https://arxiv.org/abs/1610.00006} {arXiv:1610.00006
  [physics.ins-det]} \BibitemShut {NoStop}%
\bibitem [{\citenamefont {Angloher}\ \emph {et~al.}(2023)\citenamefont
  {Angloher} \emph {et~al.}}]{COSINUS:2023kqd}%
  \BibitemOpen
  \bibfield  {author} {\bibinfo {author} {\bibfnamefont {G.}~\bibnamefont
  {Angloher}} \emph {et~al.} (\bibinfo {collaboration} {COSINUS
  Collaboration}),\ }\href@noop {} {\  (\bibinfo {year} {2023})},\ \Eprint
  {https://arxiv.org/abs/2307.11139} {arXiv:2307.11139 [astro-ph.CO]}
  \BibitemShut {NoStop}%
\bibitem [{\citenamefont {De~Gerone}\ \emph {et~al.}(2022)\citenamefont
  {De~Gerone} \emph {et~al.}}]{DeGerone:2022dxb}%
  \BibitemOpen
  \bibfield  {author} {\bibinfo {author} {\bibfnamefont {M.}~\bibnamefont
  {De~Gerone}} \emph {et~al.},\ }\href
  {https://doi.org/10.1007/s10909-022-02895-6} {\bibfield  {journal} {\bibinfo
  {journal} {J. Low Temp. Phys.}\ }\textbf {\bibinfo {volume} {209}},\ \bibinfo
  {pages} {980} (\bibinfo {year} {2022})}\BibitemShut {NoStop}%
\bibitem [{\citenamefont {Baracchini}\ \emph {et~al.}(2018)\citenamefont
  {Baracchini} \emph {et~al.}}]{PTOLEMY:2018jst}%
  \BibitemOpen
  \bibfield  {author} {\bibinfo {author} {\bibfnamefont {E.}~\bibnamefont
  {Baracchini}} \emph {et~al.} (\bibinfo {collaboration} {PTOLEMY
  Collaboration}),\ }\href@noop {} {\  (\bibinfo {year} {2018})},\ \Eprint
  {https://arxiv.org/abs/1808.01892} {arXiv:1808.01892 [physics.ins-det]}
  \BibitemShut {NoStop}%
\bibitem [{\citenamefont {Arnaud}\ \emph {et~al.}(2018)\citenamefont {Arnaud}
  \emph {et~al.}}]{EDELWEISS:2017uga}%
  \BibitemOpen
  \bibfield  {author} {\bibinfo {author} {\bibfnamefont {Q.}~\bibnamefont
  {Arnaud}} \emph {et~al.} (\bibinfo {collaboration} {EDELWEISS
  Collaboration}),\ }\href {https://doi.org/10.1103/PhysRevD.97.022003}
  {\bibfield  {journal} {\bibinfo  {journal} {Phys. Rev. D}\ }\textbf {\bibinfo
  {volume} {97}},\ \bibinfo {pages} {022003} (\bibinfo {year} {2018})},\
  \Eprint {https://arxiv.org/abs/1707.04308} {arXiv:1707.04308
  [physics.ins-det]} \BibitemShut {NoStop}%
\bibitem [{\citenamefont {Cruciani}\ \emph {et~al.}(2022)\citenamefont
  {Cruciani} \emph {et~al.}}]{Cruciani:2022mbb}%
  \BibitemOpen
  \bibfield  {author} {\bibinfo {author} {\bibfnamefont {A.}~\bibnamefont
  {Cruciani}} \emph {et~al.},\ }\href {https://doi.org/10.1063/5.0128723}
  {\bibfield  {journal} {\bibinfo  {journal} {Appl. Phys. Lett.}\ }\textbf
  {\bibinfo {volume} {121}},\ \bibinfo {pages} {213504} (\bibinfo {year}
  {2022})},\ \Eprint {https://arxiv.org/abs/2209.14806} {arXiv:2209.14806
  [physics.ins-det]} \BibitemShut {NoStop}%
\bibitem [{\citenamefont {Aja}\ \emph {et~al.}(2022)\citenamefont {Aja} \emph
  {et~al.}}]{Aja:2022csb}%
  \BibitemOpen
  \bibfield  {author} {\bibinfo {author} {\bibfnamefont {B.}~\bibnamefont
  {Aja}} \emph {et~al.},\ }\href
  {https://doi.org/10.1088/1475-7516/2022/11/044} {\bibfield  {journal}
  {\bibinfo  {journal} {JCAP}\ }\textbf {\bibinfo {volume} {11}},\ \bibinfo
  {pages} {044}},\ \Eprint {https://arxiv.org/abs/2206.02980} {arXiv:2206.02980
  [hep-ex]} \BibitemShut {NoStop}%
\bibitem [{\citenamefont {Casali}\ \emph {et~al.}(2019)\citenamefont {Casali},
  \citenamefont {Bellini}, \citenamefont {Calvo}, \citenamefont {Cardani},
  \citenamefont {Castellano}, \citenamefont {Cosmelli}, \citenamefont
  {Cruciani}, \citenamefont {Domizio}, \citenamefont {Fresch}, \citenamefont
  {Goupy}, \citenamefont {Martinez}, \citenamefont {Monfardini}, \citenamefont
  {Pettinari}, \citenamefont {le~Sueur},\ and\ \citenamefont
  {Vignati}}]{Casali2019}%
  \BibitemOpen
  \bibfield  {author} {\bibinfo {author} {\bibfnamefont {N.}~\bibnamefont
  {Casali}}, \bibinfo {author} {\bibfnamefont {F.}~\bibnamefont {Bellini}},
  \bibinfo {author} {\bibfnamefont {M.}~\bibnamefont {Calvo}}, \bibinfo
  {author} {\bibfnamefont {L.}~\bibnamefont {Cardani}}, \bibinfo {author}
  {\bibfnamefont {M.}~\bibnamefont {Castellano}}, \bibinfo {author}
  {\bibfnamefont {C.}~\bibnamefont {Cosmelli}}, \bibinfo {author}
  {\bibfnamefont {A.}~\bibnamefont {Cruciani}}, \bibinfo {author}
  {\bibfnamefont {S.~D.}\ \bibnamefont {Domizio}}, \bibinfo {author}
  {\bibfnamefont {P.}~\bibnamefont {Fresch}}, \bibinfo {author} {\bibfnamefont
  {J.}~\bibnamefont {Goupy}}, \bibinfo {author} {\bibfnamefont
  {M.}~\bibnamefont {Martinez}}, \bibinfo {author} {\bibfnamefont
  {A.}~\bibnamefont {Monfardini}}, \bibinfo {author} {\bibfnamefont
  {G.}~\bibnamefont {Pettinari}}, \bibinfo {author} {\bibfnamefont
  {H.}~\bibnamefont {le~Sueur}},\ and\ \bibinfo {author} {\bibfnamefont
  {M.}~\bibnamefont {Vignati}},\ }\href
  {https://doi.org/10.1016/j.nima.2018.10.079} {\bibfield  {journal} {\bibinfo
  {journal} {Nuclear Instruments and Methods in Physics Research Section A:
  Accelerators, Spectrometers, Detectors and Associated Equipment}\ }\textbf
  {\bibinfo {volume} {936}},\ \bibinfo {pages} {166} (\bibinfo {year}
  {2019})}\BibitemShut {NoStop}%
\bibitem [{\citenamefont {Chiles}\ \emph {et~al.}(2022)\citenamefont {Chiles}
  \emph {et~al.}}]{Chiles:2021gxk}%
  \BibitemOpen
  \bibfield  {author} {\bibinfo {author} {\bibfnamefont {J.}~\bibnamefont
  {Chiles}} \emph {et~al.},\ }\href
  {https://doi.org/10.1103/PhysRevLett.128.231802} {\bibfield  {journal}
  {\bibinfo  {journal} {Phys. Rev. Lett.}\ }\textbf {\bibinfo {volume} {128}},\
  \bibinfo {pages} {231802} (\bibinfo {year} {2022})},\ \Eprint
  {https://arxiv.org/abs/2110.01582} {arXiv:2110.01582 [hep-ex]} \BibitemShut
  {NoStop}%
\bibitem [{\citenamefont {Mantegazzini}\ \emph {et~al.}(2023)\citenamefont
  {Mantegazzini}, \citenamefont {Kovac}, \citenamefont {Enss}, \citenamefont
  {Fleischmann}, \citenamefont {Griedel},\ and\ \citenamefont
  {Gastaldo}}]{Mantegazzini:2023igy}%
  \BibitemOpen
  \bibfield  {author} {\bibinfo {author} {\bibfnamefont {F.}~\bibnamefont
  {Mantegazzini}}, \bibinfo {author} {\bibfnamefont {N.}~\bibnamefont {Kovac}},
  \bibinfo {author} {\bibfnamefont {C.}~\bibnamefont {Enss}}, \bibinfo {author}
  {\bibfnamefont {A.}~\bibnamefont {Fleischmann}}, \bibinfo {author}
  {\bibfnamefont {M.}~\bibnamefont {Griedel}},\ and\ \bibinfo {author}
  {\bibfnamefont {L.}~\bibnamefont {Gastaldo}},\ }\href
  {https://doi.org/10.1016/j.nima.2023.168564} {\bibfield  {journal} {\bibinfo
  {journal} {Nucl. Instrum. Meth. A}\ }\textbf {\bibinfo {volume} {1055}},\
  \bibinfo {pages} {168564} (\bibinfo {year} {2023})},\ \Eprint
  {https://arxiv.org/abs/2301.06455} {arXiv:2301.06455 [physics.ins-det]}
  \BibitemShut {NoStop}%
\bibitem [{\citenamefont {Irwin}\ \emph {et~al.}(1995)\citenamefont {Irwin},
  \citenamefont {Nam}, \citenamefont {Cabrera}, \citenamefont {Chugg},\ and\
  \citenamefont {Young}}]{IrwinRSI95}%
  \BibitemOpen
  \bibfield  {author} {\bibinfo {author} {\bibfnamefont {K.~D.}\ \bibnamefont
  {Irwin}}, \bibinfo {author} {\bibfnamefont {S.~W.}\ \bibnamefont {Nam}},
  \bibinfo {author} {\bibfnamefont {B.}~\bibnamefont {Cabrera}}, \bibinfo
  {author} {\bibfnamefont {B.}~\bibnamefont {Chugg}},\ and\ \bibinfo {author}
  {\bibfnamefont {B.~A.}\ \bibnamefont {Young}},\ }\href
  {https://doi.org/10.1063/1.1146105} {\bibfield  {journal} {\bibinfo
  {journal} {Rev. Sci. Instrum.}\ }\textbf {\bibinfo {volume} {66}},\ \bibinfo
  {pages} {5322} (\bibinfo {year} {1995})}\BibitemShut {NoStop}%
\bibitem [{\citenamefont {Fink}\ \emph {et~al.}(2020)\citenamefont {Fink} \emph
  {et~al.}}]{FinkAIPAdv20}%
  \BibitemOpen
  \bibfield  {author} {\bibinfo {author} {\bibfnamefont {C.~W.}\ \bibnamefont
  {Fink}} \emph {et~al.},\ }\href {https://doi.org/10.1063/5.0011130}
  {\bibfield  {journal} {\bibinfo  {journal} {AIP Adv.}\ }\textbf {\bibinfo
  {volume} {10}},\ \bibinfo {pages} {085221} (\bibinfo {year} {2020})},\
  \Eprint {https://arxiv.org/abs/2004.10257} {arXiv:2004.10257
  [physics.ins-det]} \BibitemShut {NoStop}%
\bibitem [{\citenamefont {Griffin}\ \emph {et~al.}(2020)\citenamefont
  {Griffin}, \citenamefont {Inzani}, \citenamefont {Trickle}, \citenamefont
  {Zhang},\ and\ \citenamefont {Zurek}}]{GriffinPRD20}%
  \BibitemOpen
  \bibfield  {author} {\bibinfo {author} {\bibfnamefont {S.~M.}\ \bibnamefont
  {Griffin}}, \bibinfo {author} {\bibfnamefont {K.}~\bibnamefont {Inzani}},
  \bibinfo {author} {\bibfnamefont {T.}~\bibnamefont {Trickle}}, \bibinfo
  {author} {\bibfnamefont {Z.}~\bibnamefont {Zhang}},\ and\ \bibinfo {author}
  {\bibfnamefont {K.~M.}\ \bibnamefont {Zurek}},\ }\href
  {https://doi.org/10.1103/PhysRevD.101.055004} {\bibfield  {journal} {\bibinfo
   {journal} {Phys. Rev. D}\ }\textbf {\bibinfo {volume} {101}},\ \bibinfo
  {pages} {055004} (\bibinfo {year} {2020})},\ \Eprint
  {https://arxiv.org/abs/1910.10716} {arXiv:1910.10716 [hep-ph]} \BibitemShut
  {NoStop}%
\bibitem [{\citenamefont {Fugallo}\ and\ \citenamefont
  {Colombo}(2018)}]{FugalloPS18}%
  \BibitemOpen
  \bibfield  {author} {\bibinfo {author} {\bibfnamefont {G.}~\bibnamefont
  {Fugallo}}\ and\ \bibinfo {author} {\bibfnamefont {L.}~\bibnamefont
  {Colombo}},\ }\href {https://doi.org/10.1088/1402-4896/aaa6f3} {\bibfield
  {journal} {\bibinfo  {journal} {Physica Scripta}\ }\textbf {\bibinfo {volume}
  {93}},\ \bibinfo {pages} {043002} (\bibinfo {year} {2018})}\BibitemShut
  {NoStop}%
\bibitem [{\citenamefont {Martinez}\ \emph {et~al.}(2019)\citenamefont
  {Martinez}, \citenamefont {Cardani}, \citenamefont {Casali}, \citenamefont
  {Cruciani}, \citenamefont {Pettinari},\ and\ \citenamefont
  {Vignati}}]{MartinezPRAppl19}%
  \BibitemOpen
  \bibfield  {author} {\bibinfo {author} {\bibfnamefont {M.}~\bibnamefont
  {Martinez}}, \bibinfo {author} {\bibfnamefont {L.}~\bibnamefont {Cardani}},
  \bibinfo {author} {\bibfnamefont {N.}~\bibnamefont {Casali}}, \bibinfo
  {author} {\bibfnamefont {A.}~\bibnamefont {Cruciani}}, \bibinfo {author}
  {\bibfnamefont {G.}~\bibnamefont {Pettinari}},\ and\ \bibinfo {author}
  {\bibfnamefont {M.}~\bibnamefont {Vignati}},\ }\href
  {https://doi.org/10.1103/PhysRevApplied.11.064025} {\bibfield  {journal}
  {\bibinfo  {journal} {Phys. Rev. Appl.}\ }\textbf {\bibinfo {volume} {11}},\
  \bibinfo {pages} {064025} (\bibinfo {year} {2019})},\ \Eprint
  {https://arxiv.org/abs/1805.02495} {arXiv:1805.02495 [physics.ins-det]}
  \BibitemShut {NoStop}%
\bibitem [{gea()}]{geant}%
  \BibitemOpen
  \href@noop {} {}\bibinfo {howpublished} {\url{https://geant4.web.cern.ch/}},\
  \bibinfo {note} {accessed: \today}\BibitemShut {NoStop}%
\bibitem [{\citenamefont {Cox}\ \emph {et~al.}(2019)\citenamefont {Cox},
  \citenamefont {Melia},\ and\ \citenamefont {Rajendran}}]{Cox:2019cod}%
  \BibitemOpen
  \bibfield  {author} {\bibinfo {author} {\bibfnamefont {P.}~\bibnamefont
  {Cox}}, \bibinfo {author} {\bibfnamefont {T.}~\bibnamefont {Melia}},\ and\
  \bibinfo {author} {\bibfnamefont {S.}~\bibnamefont {Rajendran}},\ }\href
  {https://doi.org/10.1103/PhysRevD.100.055011} {\bibfield  {journal} {\bibinfo
   {journal} {Phys. Rev. D}\ }\textbf {\bibinfo {volume} {100}},\ \bibinfo
  {pages} {055011} (\bibinfo {year} {2019})},\ \Eprint
  {https://arxiv.org/abs/1905.05575} {arXiv:1905.05575 [hep-ph]} \BibitemShut
  {NoStop}%
\bibitem [{\citenamefont {Trickle}\ \emph {et~al.}(2020)\citenamefont
  {Trickle}, \citenamefont {Zhang}, \citenamefont {Zurek}, \citenamefont
  {Inzani},\ and\ \citenamefont {Griffin}}]{Trickle:2019nya}%
  \BibitemOpen
  \bibfield  {author} {\bibinfo {author} {\bibfnamefont {T.}~\bibnamefont
  {Trickle}}, \bibinfo {author} {\bibfnamefont {Z.}~\bibnamefont {Zhang}},
  \bibinfo {author} {\bibfnamefont {K.~M.}\ \bibnamefont {Zurek}}, \bibinfo
  {author} {\bibfnamefont {K.}~\bibnamefont {Inzani}},\ and\ \bibinfo {author}
  {\bibfnamefont {S.~M.}\ \bibnamefont {Griffin}},\ }\href
  {https://doi.org/10.1007/JHEP03(2020)036} {\bibfield  {journal} {\bibinfo
  {journal} {JHEP}\ }\textbf {\bibinfo {volume} {03}},\ \bibinfo {pages}
  {036}},\ \Eprint {https://arxiv.org/abs/1910.08092} {arXiv:1910.08092
  [hep-ph]} \BibitemShut {NoStop}%
\bibitem [{\citenamefont {Trickle}\ \emph {et~al.}(2022)\citenamefont
  {Trickle}, \citenamefont {Zhang},\ and\ \citenamefont
  {Zurek}}]{Trickle:2020oki}%
  \BibitemOpen
  \bibfield  {author} {\bibinfo {author} {\bibfnamefont {T.}~\bibnamefont
  {Trickle}}, \bibinfo {author} {\bibfnamefont {Z.}~\bibnamefont {Zhang}},\
  and\ \bibinfo {author} {\bibfnamefont {K.~M.}\ \bibnamefont {Zurek}},\ }\href
  {https://doi.org/10.1103/PhysRevD.105.015001} {\bibfield  {journal} {\bibinfo
   {journal} {Phys. Rev. D}\ }\textbf {\bibinfo {volume} {105}},\ \bibinfo
  {pages} {015001} (\bibinfo {year} {2022})},\ \Eprint
  {https://arxiv.org/abs/2009.13534} {arXiv:2009.13534 [hep-ph]} \BibitemShut
  {NoStop}%
\bibitem [{\citenamefont {Read}(2014)}]{Read:2014qva}%
  \BibitemOpen
  \bibfield  {author} {\bibinfo {author} {\bibfnamefont {J.~I.}\ \bibnamefont
  {Read}},\ }\href {https://doi.org/10.1088/0954-3899/41/6/063101} {\bibfield
  {journal} {\bibinfo  {journal} {J. Phys. G}\ }\textbf {\bibinfo {volume}
  {41}},\ \bibinfo {pages} {063101} (\bibinfo {year} {2014})},\ \Eprint
  {https://arxiv.org/abs/1404.1938} {arXiv:1404.1938 [astro-ph.GA]}
  \BibitemShut {NoStop}%
\bibitem [{\citenamefont {de~Salas}\ and\ \citenamefont
  {Widmark}(2021)}]{deSalas:2020hbh}%
  \BibitemOpen
  \bibfield  {author} {\bibinfo {author} {\bibfnamefont {P.~F.}\ \bibnamefont
  {de~Salas}}\ and\ \bibinfo {author} {\bibfnamefont {A.}~\bibnamefont
  {Widmark}},\ }\href {https://doi.org/10.1088/1361-6633/ac24e7} {\bibfield
  {journal} {\bibinfo  {journal} {Rept. Prog. Phys.}\ }\textbf {\bibinfo
  {volume} {84}},\ \bibinfo {pages} {104901} (\bibinfo {year} {2021})},\
  \Eprint {https://arxiv.org/abs/2012.11477} {arXiv:2012.11477 [astro-ph.GA]}
  \BibitemShut {NoStop}%
\bibitem [{\citenamefont {Baxter}\ \emph {et~al.}(2021)\citenamefont {Baxter}
  \emph {et~al.}}]{Baxter:2021pqo}%
  \BibitemOpen
  \bibfield  {author} {\bibinfo {author} {\bibfnamefont {D.}~\bibnamefont
  {Baxter}} \emph {et~al.},\ }\href
  {https://doi.org/10.1140/epjc/s10052-021-09655-y} {\bibfield  {journal}
  {\bibinfo  {journal} {Eur. Phys. J. C}\ }\textbf {\bibinfo {volume} {81}},\
  \bibinfo {pages} {907} (\bibinfo {year} {2021})},\ \Eprint
  {https://arxiv.org/abs/2105.00599} {arXiv:2105.00599 [hep-ex]} \BibitemShut
  {NoStop}%
\bibitem [{Pho()}]{PhonoDark}%
  \BibitemOpen
  \href@noop {} {\bibinfo {title} {Phonodark}},\ \bibinfo {howpublished}
  {\url{https://github.com/tanner-trickle/PhonoDark}},\ \bibinfo {note}
  {accessed: \today}\BibitemShut {NoStop}%
\bibitem [{Dar()}]{DarkELF}%
  \BibitemOpen
  \href@noop {} {\bibinfo {title} {Darkelf}},\ \bibinfo {howpublished}
  {\url{https://github.com/tongylin/DarkELF}},\ \bibinfo {note} {accessed:
  \today}\BibitemShut {NoStop}%
\bibitem [{\citenamefont {Knapen}\ \emph {et~al.}(2021)\citenamefont {Knapen},
  \citenamefont {Kozaczuk},\ and\ \citenamefont {Lin}}]{Knapen:2021run}%
  \BibitemOpen
  \bibfield  {author} {\bibinfo {author} {\bibfnamefont {S.}~\bibnamefont
  {Knapen}}, \bibinfo {author} {\bibfnamefont {J.}~\bibnamefont {Kozaczuk}},\
  and\ \bibinfo {author} {\bibfnamefont {T.}~\bibnamefont {Lin}},\ }\href
  {https://doi.org/10.1103/PhysRevD.104.015031} {\bibfield  {journal} {\bibinfo
   {journal} {Phys. Rev. D}\ }\textbf {\bibinfo {volume} {104}},\ \bibinfo
  {pages} {015031} (\bibinfo {year} {2021})},\ \Eprint
  {https://arxiv.org/abs/2101.08275} {arXiv:2101.08275 [hep-ph]} \BibitemShut
  {NoStop}%
\bibitem [{\citenamefont {Knapen}\ \emph {et~al.}(2022)\citenamefont {Knapen},
  \citenamefont {Kozaczuk},\ and\ \citenamefont {Lin}}]{Knapen:2021bwg}%
  \BibitemOpen
  \bibfield  {author} {\bibinfo {author} {\bibfnamefont {S.}~\bibnamefont
  {Knapen}}, \bibinfo {author} {\bibfnamefont {J.}~\bibnamefont {Kozaczuk}},\
  and\ \bibinfo {author} {\bibfnamefont {T.}~\bibnamefont {Lin}},\ }\href
  {https://doi.org/10.1103/PhysRevD.105.015014} {\bibfield  {journal} {\bibinfo
   {journal} {Phys. Rev. D}\ }\textbf {\bibinfo {volume} {105}},\ \bibinfo
  {pages} {015014} (\bibinfo {year} {2022})},\ \Eprint
  {https://arxiv.org/abs/2104.12786} {arXiv:2104.12786 [hep-ph]} \BibitemShut
  {NoStop}%
\bibitem [{\citenamefont {Campbell-Deem}\ \emph {et~al.}(2020)\citenamefont
  {Campbell-Deem}, \citenamefont {Cox}, \citenamefont {Knapen}, \citenamefont
  {Lin},\ and\ \citenamefont {Melia}}]{Campbell-Deem:2019hdx}%
  \BibitemOpen
  \bibfield  {author} {\bibinfo {author} {\bibfnamefont {B.}~\bibnamefont
  {Campbell-Deem}}, \bibinfo {author} {\bibfnamefont {P.}~\bibnamefont {Cox}},
  \bibinfo {author} {\bibfnamefont {S.}~\bibnamefont {Knapen}}, \bibinfo
  {author} {\bibfnamefont {T.}~\bibnamefont {Lin}},\ and\ \bibinfo {author}
  {\bibfnamefont {T.}~\bibnamefont {Melia}},\ }\href
  {https://doi.org/10.1103/PhysRevD.101.036006} {\bibfield  {journal} {\bibinfo
   {journal} {Phys. Rev. D}\ }\textbf {\bibinfo {volume} {101}},\ \bibinfo
  {pages} {036006} (\bibinfo {year} {2020})},\ \bibinfo {note} {[Erratum:
  Phys.Rev.D 102, 019904 (2020)]},\ \Eprint {https://arxiv.org/abs/1911.03482}
  {arXiv:1911.03482 [hep-ph]} \BibitemShut {NoStop}%
\bibitem [{\citenamefont {Kahn}\ \emph {et~al.}(2021)\citenamefont {Kahn},
  \citenamefont {Krnjaic},\ and\ \citenamefont {Mandava}}]{Kahn:2020fef}%
  \BibitemOpen
  \bibfield  {author} {\bibinfo {author} {\bibfnamefont {Y.}~\bibnamefont
  {Kahn}}, \bibinfo {author} {\bibfnamefont {G.}~\bibnamefont {Krnjaic}},\ and\
  \bibinfo {author} {\bibfnamefont {B.}~\bibnamefont {Mandava}},\ }\href
  {https://doi.org/10.1103/PhysRevLett.127.081804} {\bibfield  {journal}
  {\bibinfo  {journal} {Phys. Rev. Lett.}\ }\textbf {\bibinfo {volume} {127}},\
  \bibinfo {pages} {081804} (\bibinfo {year} {2021})},\ \Eprint
  {https://arxiv.org/abs/2011.09477} {arXiv:2011.09477 [hep-ph]} \BibitemShut
  {NoStop}%
\bibitem [{\citenamefont {Campbell-Deem}\ \emph {et~al.}(2022)\citenamefont
  {Campbell-Deem}, \citenamefont {Knapen}, \citenamefont {Lin},\ and\
  \citenamefont {Villarama}}]{Campbell-Deem:2022fqm}%
  \BibitemOpen
  \bibfield  {author} {\bibinfo {author} {\bibfnamefont {B.}~\bibnamefont
  {Campbell-Deem}}, \bibinfo {author} {\bibfnamefont {S.}~\bibnamefont
  {Knapen}}, \bibinfo {author} {\bibfnamefont {T.}~\bibnamefont {Lin}},\ and\
  \bibinfo {author} {\bibfnamefont {E.}~\bibnamefont {Villarama}},\ }\href
  {https://doi.org/10.1103/PhysRevD.106.036019} {\bibfield  {journal} {\bibinfo
   {journal} {Phys. Rev. D}\ }\textbf {\bibinfo {volume} {106}},\ \bibinfo
  {pages} {036019} (\bibinfo {year} {2022})},\ \Eprint
  {https://arxiv.org/abs/2205.02250} {arXiv:2205.02250 [hep-ph]} \BibitemShut
  {NoStop}%
\bibitem [{mat()}]{matproj}%
  \BibitemOpen
  \href@noop {} {}\bibinfo {howpublished}
  {\url{https://materialsproject.org/}},\ \bibinfo {note} {accessed:
  \today}\BibitemShut {NoStop}%
\bibitem [{alm()}]{almadb}%
  \BibitemOpen
  \href@noop {} {}\bibinfo {howpublished}
  {\url{https://almabte.bitbucket.io/database/}},\ \bibinfo {note} {accessed:
  \today}\BibitemShut {NoStop}%
\bibitem [{pho()}]{phononkyoto}%
  \BibitemOpen
  \href@noop {} {}\bibinfo {howpublished}
  {\url{//phonondb.mtl.kyoto-u.ac.jp/index.html}},\ \bibinfo {note} {accessed:
  \today}\BibitemShut {NoStop}%
\bibitem [{\citenamefont {Schober}\ \emph {et~al.}(1993)\citenamefont
  {Schober}, \citenamefont {Strauch},\ and\ \citenamefont
  {Dorner}}]{SchoberZPB93}%
  \BibitemOpen
  \bibfield  {author} {\bibinfo {author} {\bibfnamefont {H.}~\bibnamefont
  {Schober}}, \bibinfo {author} {\bibfnamefont {D.}~\bibnamefont {Strauch}},\
  and\ \bibinfo {author} {\bibfnamefont {B.}~\bibnamefont {Dorner}},\ }\href
  {https://doi.org/10.1007/BF01308745} {\bibfield  {journal} {\bibinfo
  {journal} {Z. Physik B - Condensed Matter}\ }\textbf {\bibinfo {volume}
  {92}},\ \bibinfo {pages} {273} (\bibinfo {year} {1993})}\BibitemShut
  {NoStop}%
\bibitem [{\citenamefont {Togo}\ \emph {et~al.}(2015)\citenamefont {Togo},
  \citenamefont {Chaput},\ and\ \citenamefont {Tanaka}}]{TogoPRB15}%
  \BibitemOpen
  \bibfield  {author} {\bibinfo {author} {\bibfnamefont {A.}~\bibnamefont
  {Togo}}, \bibinfo {author} {\bibfnamefont {L.}~\bibnamefont {Chaput}},\ and\
  \bibinfo {author} {\bibfnamefont {I.}~\bibnamefont {Tanaka}},\ }\href
  {https://doi.org/10.1103/PhysRevB.91.094306} {\bibfield  {journal} {\bibinfo
  {journal} {Phys. Rev. B}\ }\textbf {\bibinfo {volume} {91}},\ \bibinfo
  {pages} {094306} (\bibinfo {year} {2015})},\ \Eprint
  {https://arxiv.org/abs/1501.00691} {arXiv:1501.00691 [cond-mat.mtrl-sci]}
  \BibitemShut {NoStop}%
\bibitem [{\citenamefont {Ziman}(1960)}]{ZimanEP}%
  \BibitemOpen
  \bibfield  {author} {\bibinfo {author} {\bibfnamefont {J.~M.}\ \bibnamefont
  {Ziman}},\ }\href@noop {} {\emph {\bibinfo {title} {Electrons and Phonons}}}\
  (\bibinfo  {publisher} {Oxford University Press},\ \bibinfo {year} {1960})\
  p.\ \bibinfo {pages} {463}\BibitemShut {NoStop}%
\bibitem [{\citenamefont {Landon}\ and\ \citenamefont
  {Hadjiconstantinou}(2014)}]{LandonJAP14}%
  \BibitemOpen
  \bibfield  {author} {\bibinfo {author} {\bibfnamefont {C.~D.}\ \bibnamefont
  {Landon}}\ and\ \bibinfo {author} {\bibfnamefont {N.~G.}\ \bibnamefont
  {Hadjiconstantinou}},\ }\href {https://doi.org/10.1063/1.4898090} {\bibfield
  {journal} {\bibinfo  {journal} {J. Appl. Phys.}\ }\textbf {\bibinfo {volume}
  {116}},\ \bibinfo {pages} {163502} (\bibinfo {year} {2014})}\BibitemShut
  {NoStop}%
\bibitem [{\citenamefont {Tamura}(1983)}]{TamuraPRB83}%
  \BibitemOpen
  \bibfield  {author} {\bibinfo {author} {\bibfnamefont {S.}~\bibnamefont
  {Tamura}},\ }\href {https://doi.org/10.1103/PhysRevB.27.858} {\bibfield
  {journal} {\bibinfo  {journal} {Phys. Rev. B}\ }\textbf {\bibinfo {volume}
  {27}},\ \bibinfo {pages} {858} (\bibinfo {year} {1983})}\BibitemShut
  {NoStop}%
\bibitem [{\citenamefont {Carrete}\ \emph {et~al.}(2017)\citenamefont
  {Carrete}, \citenamefont {Vermeersch}, \citenamefont {Katre}, \citenamefont
  {van Roekeghem}, \citenamefont {Wang}, \citenamefont {Madsen},\ and\
  \citenamefont {Mingo}}]{CarreteCPC17}%
  \BibitemOpen
  \bibfield  {author} {\bibinfo {author} {\bibfnamefont {J.}~\bibnamefont
  {Carrete}}, \bibinfo {author} {\bibfnamefont {B.}~\bibnamefont {Vermeersch}},
  \bibinfo {author} {\bibfnamefont {A.}~\bibnamefont {Katre}}, \bibinfo
  {author} {\bibfnamefont {A.}~\bibnamefont {van Roekeghem}}, \bibinfo {author}
  {\bibfnamefont {T.}~\bibnamefont {Wang}}, \bibinfo {author} {\bibfnamefont
  {G.~K.}\ \bibnamefont {Madsen}},\ and\ \bibinfo {author} {\bibfnamefont
  {N.}~\bibnamefont {Mingo}},\ }\href
  {https://doi.org/10.1016/j.cpc.2017.06.023} {\bibfield  {journal} {\bibinfo
  {journal} {Comput. Phys. Commun.}\ }\textbf {\bibinfo {volume} {220}},\
  \bibinfo {pages} {351} (\bibinfo {year} {2017})},\ \Eprint
  {https://arxiv.org/abs/1704.04142} {arXiv:1704.04142 [physics.comp-ph]}
  \BibitemShut {NoStop}%
\bibitem [{\citenamefont {Dongre}\ \emph {et~al.}(2018)\citenamefont {Dongre},
  \citenamefont {Carrete}, \citenamefont {Mingo},\ and\ \citenamefont
  {Madsen}}]{DongreMRSComm18}%
  \BibitemOpen
  \bibfield  {author} {\bibinfo {author} {\bibfnamefont {B.}~\bibnamefont
  {Dongre}}, \bibinfo {author} {\bibfnamefont {J.}~\bibnamefont {Carrete}},
  \bibinfo {author} {\bibfnamefont {N.}~\bibnamefont {Mingo}},\ and\ \bibinfo
  {author} {\bibfnamefont {G.~K.}\ \bibnamefont {Madsen}},\ }\href
  {https://doi.org/10.1557/mrc.2018.161} {\bibfield  {journal} {\bibinfo
  {journal} {MRS Commun.}\ }\textbf {\bibinfo {volume} {8}},\ \bibinfo {pages}
  {1119} (\bibinfo {year} {2018})},\ \Eprint {https://arxiv.org/abs/1806.08557}
  {arXiv:1806.08557 [cond-mat.mtrl-sci]} \BibitemShut {NoStop}%
\bibitem [{\citenamefont {Lundstrom}(2002)}]{LundstromBOOK2002}%
  \BibitemOpen
  \bibfield  {author} {\bibinfo {author} {\bibfnamefont {M.}~\bibnamefont
  {Lundstrom}},\ }\href@noop {} {\bibfield  {journal} {\bibinfo  {journal}
  {Measurement Science and Technology}\ }\textbf {\bibinfo {volume} {13}},\
  \bibinfo {pages} {230} (\bibinfo {year} {2002})}\BibitemShut {NoStop}%
\bibitem [{\citenamefont {Carrete}\ \emph {et~al.}(2019)\citenamefont
  {Carrete}, \citenamefont {L\'opez-Su\'arez}, \citenamefont {Raya-Moreno},
  \citenamefont {Bochkarev}, \citenamefont {Royo}, \citenamefont {Madsen},
  \citenamefont {Cartoix\`a}, \citenamefont {Mingo},\ and\ \citenamefont
  {Rurali}}]{CarreteNanoscale19}%
  \BibitemOpen
  \bibfield  {author} {\bibinfo {author} {\bibfnamefont {J.}~\bibnamefont
  {Carrete}}, \bibinfo {author} {\bibfnamefont {M.}~\bibnamefont
  {L\'opez-Su\'arez}}, \bibinfo {author} {\bibfnamefont {M.}~\bibnamefont
  {Raya-Moreno}}, \bibinfo {author} {\bibfnamefont {A.~S.}\ \bibnamefont
  {Bochkarev}}, \bibinfo {author} {\bibfnamefont {M.}~\bibnamefont {Royo}},
  \bibinfo {author} {\bibfnamefont {G.~K.~H.}\ \bibnamefont {Madsen}}, \bibinfo
  {author} {\bibfnamefont {X.}~\bibnamefont {Cartoix\`a}}, \bibinfo {author}
  {\bibfnamefont {N.}~\bibnamefont {Mingo}},\ and\ \bibinfo {author}
  {\bibfnamefont {R.}~\bibnamefont {Rurali}},\ }\href
  {https://doi.org/10.1039/C9NR05274G} {\bibfield  {journal} {\bibinfo
  {journal} {Nanoscale}\ }\textbf {\bibinfo {volume} {11}},\ \bibinfo {pages}
  {16007} (\bibinfo {year} {2019})}\BibitemShut {NoStop}%
\bibitem [{\citenamefont {Swartz}\ and\ \citenamefont
  {Pohl}(1989)}]{SwartzRMP89}%
  \BibitemOpen
  \bibfield  {author} {\bibinfo {author} {\bibfnamefont {E.~T.}\ \bibnamefont
  {Swartz}}\ and\ \bibinfo {author} {\bibfnamefont {R.~O.}\ \bibnamefont
  {Pohl}},\ }\href {https://doi.org/10.1103/RevModPhys.61.605} {\bibfield
  {journal} {\bibinfo  {journal} {Rev. Mod. Phys.}\ }\textbf {\bibinfo {volume}
  {61}},\ \bibinfo {pages} {605} (\bibinfo {year} {1989})}\BibitemShut
  {NoStop}%
\bibitem [{\citenamefont {Chen}\ \emph {et~al.}(2022)\citenamefont {Chen},
  \citenamefont {Xu}, \citenamefont {Zhou},\ and\ \citenamefont
  {Li}}]{ChenRMP22}%
  \BibitemOpen
  \bibfield  {author} {\bibinfo {author} {\bibfnamefont {J.}~\bibnamefont
  {Chen}}, \bibinfo {author} {\bibfnamefont {X.}~\bibnamefont {Xu}}, \bibinfo
  {author} {\bibfnamefont {J.}~\bibnamefont {Zhou}},\ and\ \bibinfo {author}
  {\bibfnamefont {B.}~\bibnamefont {Li}},\ }\href
  {https://doi.org/10.1103/RevModPhys.94.025002} {\bibfield  {journal}
  {\bibinfo  {journal} {Rev. Mod. Phys.}\ }\textbf {\bibinfo {volume} {94}},\
  \bibinfo {pages} {025002} (\bibinfo {year} {2022})}\BibitemShut {NoStop}%
\bibitem [{\citenamefont {Reddy}\ \emph {et~al.}(2005)\citenamefont {Reddy},
  \citenamefont {Castelino},\ and\ \citenamefont {Majumdar}}]{ReddyAPL05}%
  \BibitemOpen
  \bibfield  {author} {\bibinfo {author} {\bibfnamefont {P.}~\bibnamefont
  {Reddy}}, \bibinfo {author} {\bibfnamefont {K.}~\bibnamefont {Castelino}},\
  and\ \bibinfo {author} {\bibfnamefont {A.}~\bibnamefont {Majumdar}},\ }\href
  {https://doi.org/10.1063/1.2133890} {\bibfield  {journal} {\bibinfo
  {journal} {Appl. Phys. Lett.}\ }\textbf {\bibinfo {volume} {87}},\ \bibinfo
  {pages} {211908} (\bibinfo {year} {2005})}\BibitemShut {NoStop}%
\bibitem [{\citenamefont {Larroque}\ \emph {et~al.}(2018)\citenamefont
  {Larroque}, \citenamefont {Dollfus},\ and\ \citenamefont
  {Saint-Martin}}]{LarroqueJAP18}%
  \BibitemOpen
  \bibfield  {author} {\bibinfo {author} {\bibfnamefont {J.}~\bibnamefont
  {Larroque}}, \bibinfo {author} {\bibfnamefont {P.}~\bibnamefont {Dollfus}},\
  and\ \bibinfo {author} {\bibfnamefont {J.}~\bibnamefont {Saint-Martin}},\
  }\href {https://doi.org/10.1063/1.5007034} {\bibfield  {journal} {\bibinfo
  {journal} {J. Appl. Phys.}\ }\textbf {\bibinfo {volume} {123}},\ \bibinfo
  {pages} {025702} (\bibinfo {year} {2018})}\BibitemShut {NoStop}%
\bibitem [{Note1()}]{Note1}%
  \BibitemOpen
  \bibinfo {note} {That is, the perpendicular momentum of the outgoing phonon
  is fully randomized under the energy conservation constraint. This is in
  contrast with specular scattering in which the outgoing momentum is uniquely
  determined by the input phonon.}\BibitemShut {Stop}%
\bibitem [{\citenamefont {Ward}\ \emph {et~al.}(2009)\citenamefont {Ward},
  \citenamefont {Broido}, \citenamefont {Stewart},\ and\ \citenamefont
  {Deinzer}}]{WardPRB09}%
  \BibitemOpen
  \bibfield  {author} {\bibinfo {author} {\bibfnamefont {A.}~\bibnamefont
  {Ward}}, \bibinfo {author} {\bibfnamefont {D.~A.}\ \bibnamefont {Broido}},
  \bibinfo {author} {\bibfnamefont {D.~A.}\ \bibnamefont {Stewart}},\ and\
  \bibinfo {author} {\bibfnamefont {G.}~\bibnamefont {Deinzer}},\ }\href
  {https://doi.org/10.1103/PhysRevB.80.125203} {\bibfield  {journal} {\bibinfo
  {journal} {Phys. Rev. B}\ }\textbf {\bibinfo {volume} {80}},\ \bibinfo
  {pages} {125203} (\bibinfo {year} {2009})}\BibitemShut {NoStop}%
\bibitem [{\citenamefont {Cepellotti}\ \emph {et~al.}(2020)\citenamefont
  {Cepellotti}, \citenamefont {Fugallo}, \citenamefont {Paulatto},
  \citenamefont {Lazzeri}, \citenamefont {Mauri},\ and\ \citenamefont
  {Marzari}}]{CepellottiNatComm15}%
  \BibitemOpen
  \bibfield  {author} {\bibinfo {author} {\bibfnamefont {A.}~\bibnamefont
  {Cepellotti}}, \bibinfo {author} {\bibfnamefont {G.}~\bibnamefont {Fugallo}},
  \bibinfo {author} {\bibfnamefont {L.}~\bibnamefont {Paulatto}}, \bibinfo
  {author} {\bibfnamefont {M.}~\bibnamefont {Lazzeri}}, \bibinfo {author}
  {\bibfnamefont {F.}~\bibnamefont {Mauri}},\ and\ \bibinfo {author}
  {\bibfnamefont {N.}~\bibnamefont {Marzari}},\ }\href
  {https://doi.org/10.1038/ncomms7400} {\bibfield  {journal} {\bibinfo
  {journal} {Nat. Commun.}\ }\textbf {\bibinfo {volume} {6}},\ \bibinfo {pages}
  {6400} (\bibinfo {year} {2020})}\BibitemShut {NoStop}%
\bibitem [{\citenamefont {Lindsay}\ \emph {et~al.}(2019)\citenamefont
  {Lindsay}, \citenamefont {Katre}, \citenamefont {Cepellotti},\ and\
  \citenamefont {Mingo}}]{LindsayJAP19}%
  \BibitemOpen
  \bibfield  {author} {\bibinfo {author} {\bibfnamefont {L.}~\bibnamefont
  {Lindsay}}, \bibinfo {author} {\bibfnamefont {A.}~\bibnamefont {Katre}},
  \bibinfo {author} {\bibfnamefont {A.}~\bibnamefont {Cepellotti}},\ and\
  \bibinfo {author} {\bibfnamefont {N.}~\bibnamefont {Mingo}},\ }\href
  {https://doi.org/10.1063/1.5108651} {\bibfield  {journal} {\bibinfo
  {journal} {J. Appl. Phys.}\ }\textbf {\bibinfo {volume} {126}},\ \bibinfo
  {pages} {050902} (\bibinfo {year} {2019})}\BibitemShut {NoStop}%
\bibitem [{\citenamefont {Li}\ \emph {et~al.}(2014)\citenamefont {Li},
  \citenamefont {Carrete}, \citenamefont {Katcho},\ and\ \citenamefont
  {Mingo}}]{LiCPC14}%
  \BibitemOpen
  \bibfield  {author} {\bibinfo {author} {\bibfnamefont {W.}~\bibnamefont
  {Li}}, \bibinfo {author} {\bibfnamefont {J.}~\bibnamefont {Carrete}},
  \bibinfo {author} {\bibfnamefont {N.~A.}\ \bibnamefont {Katcho}},\ and\
  \bibinfo {author} {\bibfnamefont {N.}~\bibnamefont {Mingo}},\ }\href
  {https://doi.org/10.1016/j.cpc.2014.02.015} {\bibfield  {journal} {\bibinfo
  {journal} {Comp. Phys. Commun.}\ }\textbf {\bibinfo {volume} {185}},\
  \bibinfo {pages} {1747} (\bibinfo {year} {2014})}\BibitemShut {NoStop}%
\bibitem [{\citenamefont {Harrelson}\ \emph {et~al.}(2021)\citenamefont
  {Harrelson}, \citenamefont {Hajar}, \citenamefont {Ashour},\ and\
  \citenamefont {Griffin}}]{HarrelsonArXiv21}%
  \BibitemOpen
  \bibfield  {author} {\bibinfo {author} {\bibfnamefont {T.~F.}\ \bibnamefont
  {Harrelson}}, \bibinfo {author} {\bibfnamefont {I.}~\bibnamefont {Hajar}},
  \bibinfo {author} {\bibfnamefont {O.~A.}\ \bibnamefont {Ashour}},\ and\
  \bibinfo {author} {\bibfnamefont {S.~M.}\ \bibnamefont {Griffin}},\
  }\href@noop {} {\  (\bibinfo {year} {2021})},\ \Eprint
  {https://arxiv.org/abs/2109.10988} {arXiv:2109.10988 [cond-mat.mtrl-sci]}
  \BibitemShut {NoStop}%
\bibitem [{\citenamefont {Fink}\ \emph {et~al.}(2021)\citenamefont {Fink} \emph
  {et~al.}}]{FinkAPL21}%
  \BibitemOpen
  \bibfield  {author} {\bibinfo {author} {\bibfnamefont {C.~W.}\ \bibnamefont
  {Fink}} \emph {et~al.} (\bibinfo {collaboration} {CPD Collaboration}),\
  }\href {https://doi.org/10.1063/5.0032372} {\bibfield  {journal} {\bibinfo
  {journal} {Appl. Phys. Lett.}\ }\textbf {\bibinfo {volume} {118}},\ \bibinfo
  {pages} {022601} (\bibinfo {year} {2021})},\ \Eprint
  {https://arxiv.org/abs/2009.14302} {arXiv:2009.14302 [physics.ins-det]}
  \BibitemShut {NoStop}%
\bibitem [{\citenamefont {Khosropanah}\ \emph {et~al.}(2010)\citenamefont
  {Khosropanah}, \citenamefont {Dirks}, \citenamefont {Parra-Border{\'i}as},
  \citenamefont {Ridder}, \citenamefont {Hijmering}, \citenamefont {van~der
  Kuur}, \citenamefont {Gottardi}, \citenamefont {Bruijn}, \citenamefont
  {Popescu}, \citenamefont {Gao},\ and\ \citenamefont
  {Hoevers}}]{Khosropanah10}%
  \BibitemOpen
  \bibfield  {author} {\bibinfo {author} {\bibfnamefont {P.}~\bibnamefont
  {Khosropanah}}, \bibinfo {author} {\bibfnamefont {B.}~\bibnamefont {Dirks}},
  \bibinfo {author} {\bibfnamefont {M.}~\bibnamefont {Parra-Border{\'i}as}},
  \bibinfo {author} {\bibfnamefont {M.}~\bibnamefont {Ridder}}, \bibinfo
  {author} {\bibfnamefont {R.}~\bibnamefont {Hijmering}}, \bibinfo {author}
  {\bibfnamefont {J.}~\bibnamefont {van~der Kuur}}, \bibinfo {author}
  {\bibfnamefont {L.}~\bibnamefont {Gottardi}}, \bibinfo {author}
  {\bibfnamefont {M.}~\bibnamefont {Bruijn}}, \bibinfo {author} {\bibfnamefont
  {M.}~\bibnamefont {Popescu}}, \bibinfo {author} {\bibfnamefont {J.~R.}\
  \bibnamefont {Gao}},\ and\ \bibinfo {author} {\bibfnamefont {H.}~\bibnamefont
  {Hoevers}},\ }in\ \href@noop {} {\emph {\bibinfo {booktitle} {Millimeter,
  Submillimeter, and Far-Infrared Detectors and Instrumentation for Astronomy
  V}}},\ Vol.\ \bibinfo {volume} {7741},\ \bibinfo {editor} {edited by\
  \bibinfo {editor} {\bibfnamefont {W.~S.}\ \bibnamefont {Holland}}\ and\
  \bibinfo {editor} {\bibfnamefont {J.}~\bibnamefont {Zmuidzinas}}},\ \bibinfo
  {organization} {International Society for Optics and Photonics}\ (\bibinfo
  {publisher} {SPIE},\ \bibinfo {year} {2010})\ p.\ \bibinfo {pages}
  {77410L}\BibitemShut {NoStop}%
\bibitem [{\citenamefont {Paolucci}\ \emph {et~al.}(2020)\citenamefont
  {Paolucci} \emph {et~al.}}]{PaolucciJAP20}%
  \BibitemOpen
  \bibfield  {author} {\bibinfo {author} {\bibfnamefont {F.}~\bibnamefont
  {Paolucci}} \emph {et~al.},\ }\href {https://doi.org/10.1063/5.0021996}
  {\bibfield  {journal} {\bibinfo  {journal} {J. Appl. Phys.}\ }\textbf
  {\bibinfo {volume} {128}},\ \bibinfo {pages} {194502} (\bibinfo {year}
  {2020})},\ \Eprint {https://arxiv.org/abs/2007.08320} {arXiv:2007.08320
  [cond-mat.mes-hall]} \BibitemShut {NoStop}%
\bibitem [{\citenamefont {Griffin}\ \emph {et~al.}(2021)\citenamefont
  {Griffin}, \citenamefont {Hochberg}, \citenamefont {Inzani}, \citenamefont
  {Kurinsky}, \citenamefont {Lin},\ and\ \citenamefont
  {Chin}}]{Griffin:2020lgd}%
  \BibitemOpen
  \bibfield  {author} {\bibinfo {author} {\bibfnamefont {S.~M.}\ \bibnamefont
  {Griffin}}, \bibinfo {author} {\bibfnamefont {Y.}~\bibnamefont {Hochberg}},
  \bibinfo {author} {\bibfnamefont {K.}~\bibnamefont {Inzani}}, \bibinfo
  {author} {\bibfnamefont {N.}~\bibnamefont {Kurinsky}}, \bibinfo {author}
  {\bibfnamefont {T.}~\bibnamefont {Lin}},\ and\ \bibinfo {author}
  {\bibfnamefont {T.}~\bibnamefont {Chin}},\ }\href
  {https://doi.org/10.1103/PhysRevD.103.075002} {\bibfield  {journal} {\bibinfo
   {journal} {Phys. Rev. D}\ }\textbf {\bibinfo {volume} {103}},\ \bibinfo
  {pages} {075002} (\bibinfo {year} {2021})},\ \Eprint
  {https://arxiv.org/abs/2008.08560} {arXiv:2008.08560 [hep-ph]} \BibitemShut
  {NoStop}%
\bibitem [{\citenamefont {Coskuner}\ \emph {et~al.}(2022)\citenamefont
  {Coskuner}, \citenamefont {Trickle}, \citenamefont {Zhang},\ and\
  \citenamefont {Zurek}}]{Coskuner:2021qxo}%
  \BibitemOpen
  \bibfield  {author} {\bibinfo {author} {\bibfnamefont {A.}~\bibnamefont
  {Coskuner}}, \bibinfo {author} {\bibfnamefont {T.}~\bibnamefont {Trickle}},
  \bibinfo {author} {\bibfnamefont {Z.}~\bibnamefont {Zhang}},\ and\ \bibinfo
  {author} {\bibfnamefont {K.~M.}\ \bibnamefont {Zurek}},\ }\href
  {https://doi.org/10.1103/PhysRevD.105.015010} {\bibfield  {journal} {\bibinfo
   {journal} {Phys. Rev. D}\ }\textbf {\bibinfo {volume} {105}},\ \bibinfo
  {pages} {015010} (\bibinfo {year} {2022})},\ \Eprint
  {https://arxiv.org/abs/2102.09567} {arXiv:2102.09567 [hep-ph]} \BibitemShut
  {NoStop}%
\bibitem [{boo()}]{boostodeint}%
  \BibitemOpen
  \href@noop {} {\bibinfo {title} {Boost.numeric.odeint}},\ \bibinfo
  {howpublished}
  {\url{https://www.boost.org/doc/libs/1_82_0/libs/numeric/odeint/doc/html/index.html}},\
  \bibinfo {note} {[Online; accessed 9-November-2023]}\BibitemShut {NoStop}%
\bibitem [{\citenamefont {Raya-Moreno}\ \emph {et~al.}(2022)\citenamefont
  {Raya-Moreno}, \citenamefont {Cartoixà},\ and\ \citenamefont
  {Carrete}}]{Raya-MorenoCPC22}%
  \BibitemOpen
  \bibfield  {author} {\bibinfo {author} {\bibfnamefont {M.}~\bibnamefont
  {Raya-Moreno}}, \bibinfo {author} {\bibfnamefont {X.}~\bibnamefont
  {Cartoixà}},\ and\ \bibinfo {author} {\bibfnamefont {J.}~\bibnamefont
  {Carrete}},\ }\href
  {https://doi.org/https://doi.org/10.1016/j.cpc.2022.108504} {\bibfield
  {journal} {\bibinfo  {journal} {Comput. Phys. Commun.}\ }\textbf {\bibinfo
  {volume} {281}},\ \bibinfo {pages} {108504} (\bibinfo {year}
  {2022})}\BibitemShut {NoStop}%
\bibitem [{\citenamefont {Landon}(2014)}]{LandonPHDTHESIS14}%
  \BibitemOpen
  \bibfield  {author} {\bibinfo {author} {\bibfnamefont {C.~D.}\ \bibnamefont
  {Landon}},\ }\emph {\bibinfo {title} {A deviational Monte Carlo formulation
  of ab initio phonon transport and its application to the study of kinetic
  effects in graphene ribbons}},\ \href
  {https://doi.org/http://hdl.handle.net/1721.1/92161} {Ph.D. thesis},\
  \bibinfo  {school} {Massachusetts Institute of Technology}, \bibinfo
  {address} {Cambridge, MA} (\bibinfo {year} {2014})\BibitemShut {NoStop}%
\end{thebibliography}


\end{document}